\documentclass[%
 aip, pof,
 amsmath,amssymb,
 reprint,%
]{revtex4-1}
\usepackage{graphicx} % Required for inserting images
\usepackage{multirow}
\usepackage[utf8]{inputenc}
\usepackage[T1]{fontenc}
\usepackage{mathptmx}
\usepackage{etoolbox}

\makeatletter
\def\@email#1#2{%
 \endgroup
 \patchcmd{\titleblock@produce}
  {\frontmatter@RRAPformat}
  {\frontmatter@RRAPformat{\produce@RRAP{*#1\href{mailto:#2}{#2}}}\frontmatter@RRAPformat}
  {}{}
}%
\makeatother

\def\ii{\mathbf{i}}
\def\j{\mathbf{j}}
\def\e{\varepsilon}
\def\s{s}

\begin{document}

\preprint{AIP/123-QED}

\title[Coexisting multiphase and interfacial behaviour of ouzo]{Coexisting multiphase and interfacial behaviour of ouzo}
% Force line breaks with \\
\author{David~N.~Sibley}
\affiliation{ 
Department of Mathematical Sciences and Interdisciplinary Centre for Mathematical Modelling, Loughborough University, Loughborough LE11 3TU, UK%\\This line break forced with \textbackslash\textbackslash
}%
\author{Benjamin~D.~Goddard}
\affiliation{ 
School of Mathematics and the Maxwell Institute for Mathematical Sciences, University of Edinburgh, Edinburgh EH9 3FD, UK%\\This line break forced with \textbackslash\textbackslash
}%
\author{Fouzia~F.~Ouali}
\affiliation{ 
Department of Physics and Mathematics, Nottingham Trent University, Clifton Lane, Nottingham NG11 8NS, UK%\\This line break forced with \textbackslash\textbackslash
}%
\author{David~J.~Fairhurst}
\affiliation{ 
School of Physics and Astronomy, University of Edinburgh, Edinburgh EH9 3FD, UK%\\This line break forced with \textbackslash\textbackslash
}%
\author{Andrew~J.~Archer}
\affiliation{ 
Department of Mathematical Sciences and Interdisciplinary Centre for Mathematical Modelling, Loughborough University, Loughborough LE11 3TU, UK%\\This line break forced with \textbackslash\textbackslash
}%

\date{\today}% It is always \today, today,

\begin{abstract}
Multi-component liquid mixtures can be both complex and fascinating, with some systems being amenable to simple experimentation at home, giving valuable insight into fundamental aspects of bulk and interfacial phase behaviour.
One particularly interesting mixture is the popular drink ouzo, which has charmed both the general public and scientists by virtue of its ability to display spontaneous emulsification when water is added.
When these two clear (and potable) liquids are poured into each other, a single milky-coloured liquid is formed.
In previous work \emph{[Archer et al., Soft Matter {\bf 20}, 5889 (2024)]}, the equilibrium phase-diagram for the stable liquid phases of ouzo was captured via experiment and modelling.
Here we consider the case when the two liquid phases also coexist with the vapour phase (i.e.~along a line of triple points) and within our model uncover the complex bulk phase behaviour for this simple beverage.
As a consequence, this leads to some interesting observations, that also apply more widely, about visualising phase diagrams in ternary systems of this type.
We also examine the interfacial behaviour, connecting microscopic density functional theory results with macroscopic (Neumann) predictions for the shape of droplets at interfaces.
\end{abstract}

\maketitle

\section{Introduction}

\begin{figure}[t]
    \centering
(a)\hspace{7.5cm}\vphantom{a}\\\includegraphics[width=0.9\linewidth]{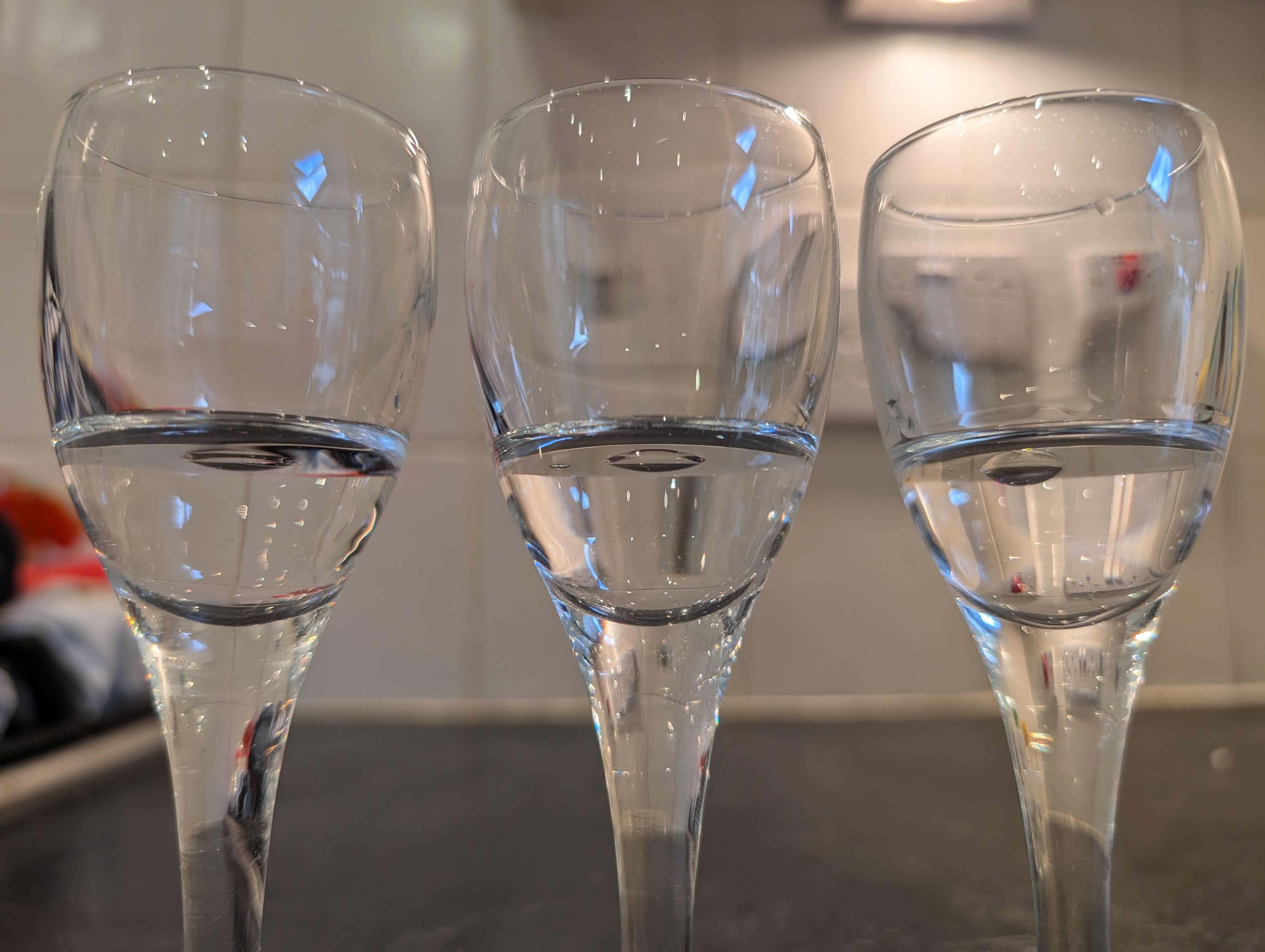}\\(b)\hspace{7.5cm}\vphantom{a}\\\includegraphics[width=0.9\linewidth]{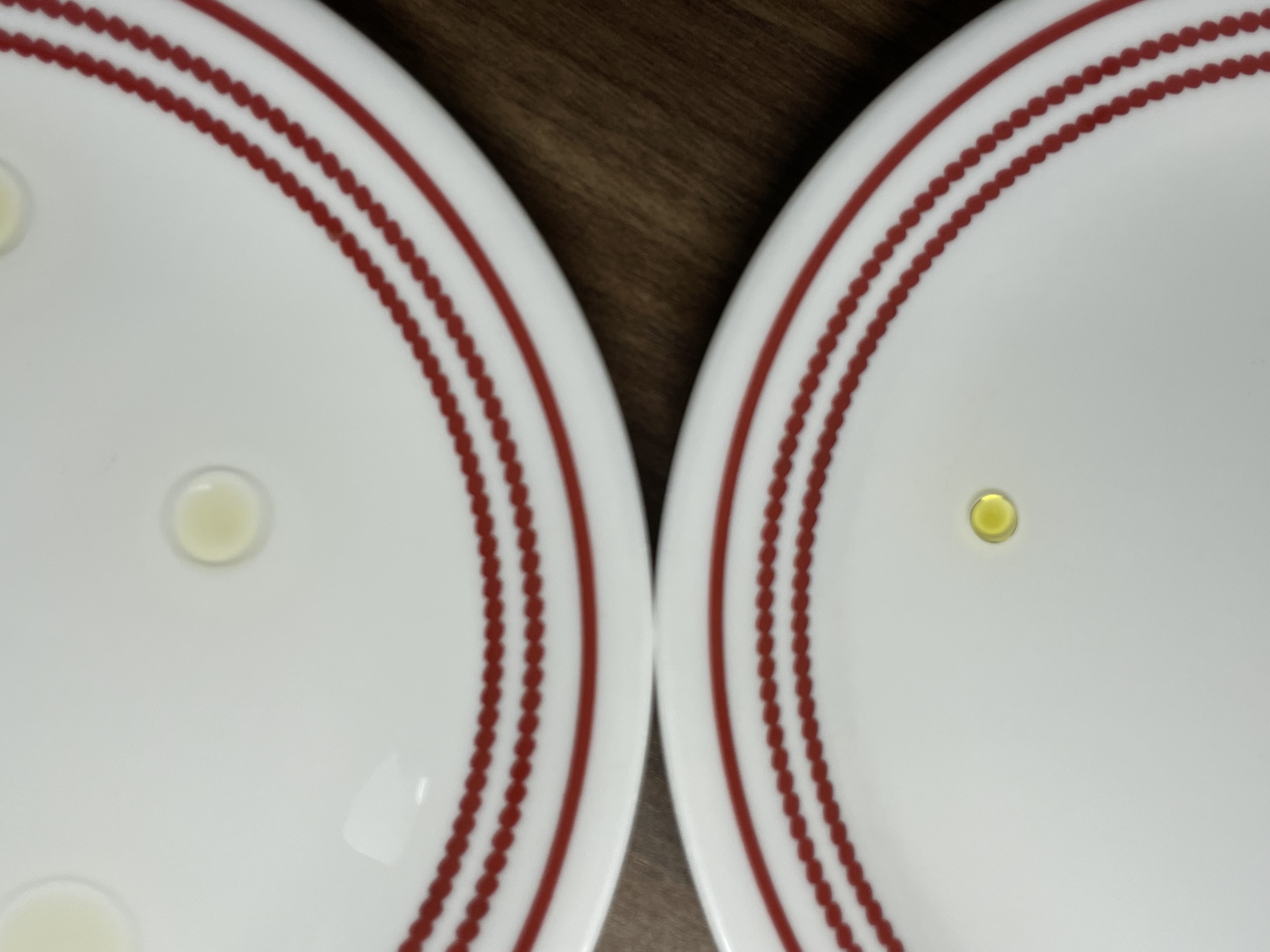}\\(c)\hspace{7.5cm}\vphantom{a}\\\includegraphics[width=0.9\linewidth]{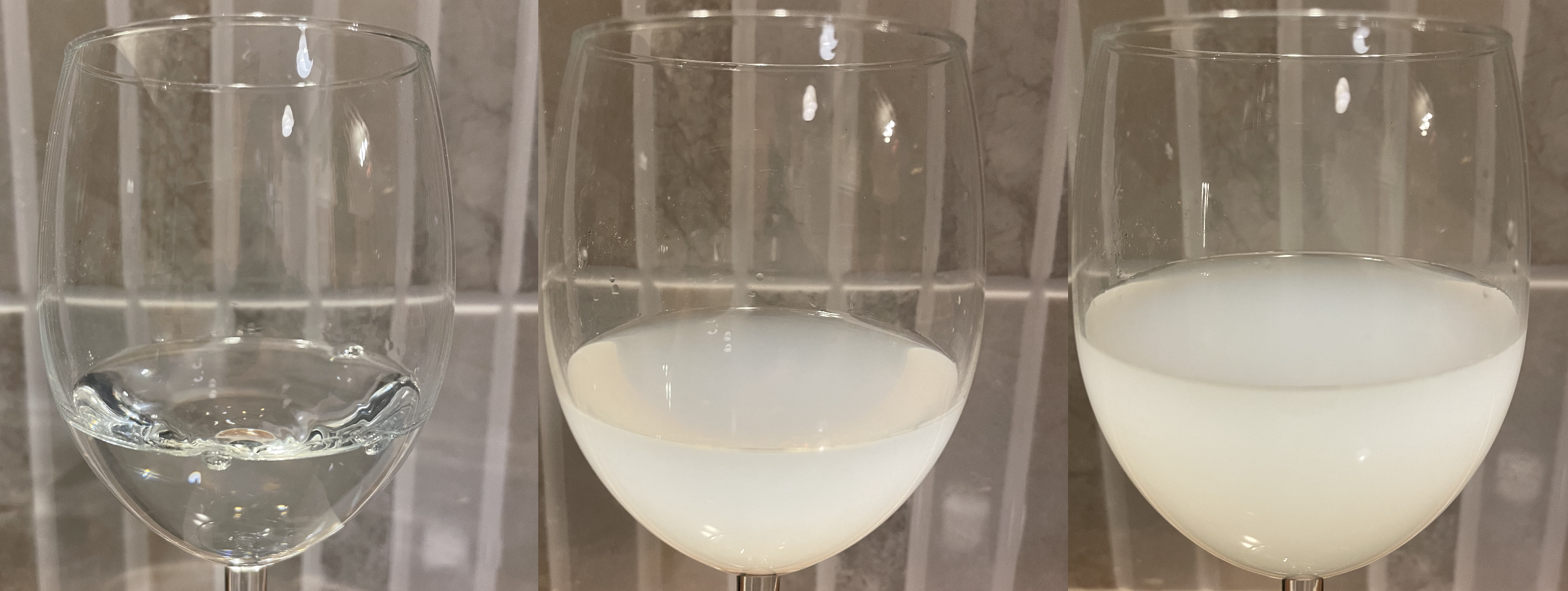}
    \caption{Some kitchen experiments with various potable/edible liquids. Images (a) and (b) show droplets on the surface of other liquids: (a) anise oil droplets of roughly the same volume on top of 100\% water (left), 50\% water 50\% gin (middle), 100\% gin (right). (b) Droplets of roughly the same volume of olive oil on water (left) and on 100\% ouzo (right). In (c) we show the ouzo effect: starting with pure ouzo on the left, moving to the right we add increasing amounts of water. 
    }
    \label{fig:kitchendrops}
\end{figure}

Immiscible liquids can exhibit behaviour which is both beautiful and educational.
One simple experiment that can be done in a kitchen is to study the spreading of oil on water, or on the surface of other liquids. Indeed, a number of the authors have done precisely this in their kitchens, see Fig.~\ref{fig:kitchendrops}(a) and (b).
Theoretical investigations of this and the underpinning bulk fluid phase behaviour can also be done in the home environment, by making sure the necessary computations can be done on a personal laptop. Here, we combine both approaches.

A small drop of oil on a liquid such as water, or some other liquid with which the oil is immiscible, forms a characteristic lens shape, with the angles of the surfaces at the contact line determined by the Neumann triangle \cite{rowlinson1982molecular}.
This experiment is relatively easy to do, although sometimes challenging to image due to the meniscus at the container's wall, and is an instructive route to understanding liquid interfacial tensions.
Moreover, understanding the properties of oil lenses on water has applications in oil-spill clean-ups, and other such applications \cite{nikolov2017oil}.

A particularly fascinating household liquid is the alcoholic drink ouzo. Ouzo is essentially made of three components: water, ethanol (roughly 60\% and 40\%, respectively) and a small amount of trans-anethole that gives the drink its aniseed flavour (alongside some other botanicals). A key reason for why it can fascinate and delight (beyond its primary purpose as an enjoyable alcoholic beverage), is that ouzo from the bottle is a clear liquid and when water (also clear) is added, the combination spontaneously turns milky, as shown in Fig.~\ref{fig:kitchendrops}(c).

In our previous work \cite{archer2024experimental}, we focused on the phase diagram for the liquid phases formed by the ouzo constituent components, to begin to better understand the formation of (seemingly stable) microscopic oil-rich droplets that visually cause the milky ouzo effect \cite{Vitale2003}.
This itself can have important uses \cite{Botet_2012,Pengetal14}, such as in the preparation of polymer nanoparticles for drug delivery \cite{lepeltier2014nanoprecipitation, goubault2020ouzo, kempe2022ouzo}, supraparticle assembly \cite{tan2019porous} and microdroplet nucleation \cite{Lohse16}.
However, it is also interesting to study the coexisting vapour phase, both from the point of view of understanding the fundamental multiphase behaviour of such a three-component system, but also as the vapour phase of alcohol products (and other volatile liquids), known as the head-space, is of significant interest for measurement, analysis and flavour perception \cite{SWIFT2023100417}.

Under a finite but reasonably wide range of temperature and pressure conditions, the three components (alcohol, oil and water) combine into three distinct and coexisting phases: an oil-rich liquid phase, a water-rich liquid phase, and a vapour phase.
Of course, at other temperatures and pressures, there are also several solid phases, but we do not consider those here.
The three-phase coexistence of the fluid phases in the phase diagram corresponds to a line of triple points \cite{rowlinson1982molecular}.
All three phases contain all three components, but in varying proportions.
Given the existence of these three stable phases, it is possible to set up physical situations where they are all present.
One exemplar where all three phases are in contact is that of an oil-rich liquid droplet sitting on a bath of the water-rich liquid phase with the vapour phase above.
There is then a three-phase contact line around the droplet.
The angles that each phase equilibrate to with each other depend on their respective surface energies/tensions and are termed Neumann angles \cite{rowlinson1982molecular}.
In this work we probe these coexisting phases and their interactions to be able to predict these surface tensions and angles from a conceptually simple mathematical model based on classical density functional theory (DFT) \cite{hansen2013theory} applied on a lattice.
One of the appealing features of DFT is that it enables one to relate the macroscopic thermodynamic and interfacial behaviour to the nature of the microscopic interactions between the constituent molecules.

\begin{figure*}
    \centering
   \includegraphics[width=0.99\linewidth]{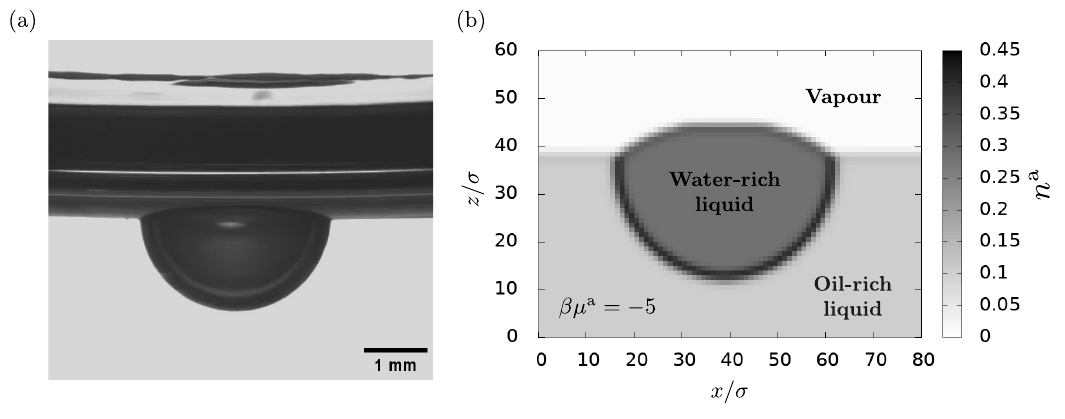}
    \caption{Comparison between experiment and modelling. For details of the modelling, see Sec.~\ref{sec:lattice}, below. (a) A photograph of a droplet containing 71.8\% water, 28.1\% ethanol, and 0.1\% trans-anethole, floating on the surface of what was pure trans-anethole, before the droplet was deposited. The image was taken 3 mins after deposition. Note the horizontal band above the droplet is the meniscus at the near-side of the cuvette. (b) The density distribution of the alcohol obtained from our lattice DFT, corresponding to a water-rich liquid droplet deposited on a bath of oil-rich liquid, with a vapour phase above. Note the highest alcohol density, indicated by the dark colouring (see grey-scale colour-bar), occurs at the interface between the bulk oil and the water rich droplet interface. This is calculated in 2D, with a fixed amount of water and oil in the system, but with alcohol chemical potential $\beta\mu^{\textrm a}=-5$ (corresponding to the alcohol number fraction within the bulk of the droplet being 28\%). Axes denote number of lattice sites and the grey shading shows the number density of the alcohol. In the experiment, the phases are not perfectly equilibrated, unlike in the DFT, which is at equilibrium, near to the bulk triple-point phase coexistence.
    }
    \label{fig:initialcomp}
\end{figure*}

Whilst our previous work already demonstrated the success of this lattice DFT model\cite{archer2024experimental}, in the spirit of kitchen-level comparisons we show in Fig.~\ref{fig:initialcomp} an example of good qualitative agreement between experiment and results from our model for floating-droplet situations.
Figure~\ref{fig:initialcomp}(a) shows an image from the lab of an ouzo-like droplet sitting on a bath of pure trans-anethole (oil).
The vapour phase is not controlled, and is mostly air.
Whilst this is clearly not a totally equilibrated situation, our equilibrium model result in Fig.~\ref{fig:initialcomp}(b), of a water-rich liquid droplet sitting on a bath of oil-rich liquid and a vapour phase of only the three ouzo components, still does a surprisingly good job of reproducing the droplet shape.

We discuss the modelling approach in detail below, but note that the majority of the detailed analysis here focuses on the situation where all three phases simultaneously coexist, i.e.~they are at a triple point. This is thus not a complete analysis of all possible multiphase coexistence---indeed the model result in Fig.~\ref{fig:initialcomp}(b) is at equilibrium but not exactly at triple-point coexistence due to the finite size of the droplet, which stems from the total number of oil and water molecules in the system being imposed as part of the calculation.
It does however uncover the essential behaviours of this coexisting system whilst providing a sensible restriction to analyse what is already a surprisingly complex and unwieldy array of phase behaviours.

This complexity is already present in the bulk fluid phase behaviour. Below, we determine the ternary phase diagram along the line of triple points, showing the lines of coexistence of the two liquids and the vapour.
We find that just displaying the regions of stability and instability within the phase diagram in an understandable way is a non-trivial exercise.
We are by no-means the first to come up against the deficiency of the classic ways of visualising ternary systems in the triangular ternary phase diagram form, but possibly for ouzo.
Our work is related to a very interesting body of biologically-related work where ternary phase diagrams are used, such as for lipids in Ref.~\onlinecite{bezlyepkina2013phase} (Fig.~5 therein has an interesting binodal and tie-line geometry, for instance) and Ref.~\onlinecite{Hirst11} (which uses a clever 3D extension to the triangle ternary phase diagram in Fig.~4 there). There are also some phase diagrams in Ref.~\onlinecite{RANA2022100944} that in some cases are similar to those we find, which is perhaps not that surprising, since it is for a similar system of ethanol-water-ethyl acetate, although using very different modelling approaches\cite{henri,abrams}. Similar calculations (but not visualised similarly) for that system are also performed elsewhere, e.g.~in Ref.~\onlinecite{KOSUGE200547}. A wide range of distilled beverages analysed using these methods have been reviewed by Puentes et al.~\cite{Puentes}

As already alluded to, our focus in this work is firstly on the extension of our lattice DFT model for the ouzo system to consider also the vapour phase and its coexistence along the triple point line with the two (oil-rich and water-rich) liquid phases. Our second goal is to then apply our microscopic DFT to determine the shape of droplets at interfaces, connect with macroscopic (Neumann) interfacial thermodynamics and thereby to shed light on our kitchen experiments. 

This paper is thus structured as follows: In Sec.~\ref{sec:2} we recall a few of the key ideas relating to the thermodynamics of liquid drops floating on liquid interfaces. Then, in Sec.~\ref{sec:lattice} we give a brief overview of our lattice DFT model for the ternary ouzo system.
In Sec.~\ref{sec:4} we present our results for the bulk phase behaviour along the triple point line, spinodals and various other relevant thermodynamic quantities, such as chemical potentials and the pressure.
Since our model is able to describe the vapour phase, we discuss properties related to this and some of the shortcomings of the traditional triangular phase diagram representation for ternary mixtures when the vapour phase is present.
We also present results for the inhomogeneous liquid, including calculating the interfacial tensions for the three interfaces between the three pairs of coexisting phases.
We also show how to calculate the density profiles for droplets of the oil-rich liquid at the interface between the water-rich liquid and its vapour.
From both the surface tensions or directly from the droplet profiles we obtain the Neumann angles between the interfaces at the three-phase contact line, so connecting our modelling to our kitchen experiments.
Finally, in Sec.~\ref{sec:5} we discuss our results and draw our conclusions, which includes a discussion of further connections to experiments.

\section{Surface modelling fundamentals}
\label{sec:2}

To set the scene, we provide a reminder of some fundamentals of the thermodynamics of liquids at interfaces.
An interface between two phases has an excess free energy that depends on the different inter-molecular forces. Essentially, by putting one bulk phase in contact with another, molecular interactions/bonds that would be present in bulk are either removed entirely or replaced by different ones (depending on the phases in contact). Also, as well as these energetic contributions, the molecular ordering at the interfaces changes, leading to entropic contributions to the interfacial tensions \cite{rowlinson1982molecular, hansen2013theory}.
Treated in the grand canonical ensemble, an equilibrium state corresponds to a minimum of the grand potential of the system, $\Omega$.
For a one component system, a given state point in this ensemble has fixed chemical potential $\mu$, system volume $V$ and temperature $T$, while the number of particles $N$, pressure $P$ and energy of the system are free to vary and adapt.
For the ternary system considered here, there are three chemical potential values, $\mu^{\textrm{a}}$, $\mu^{\textrm{o}}$ and $\mu^{\textrm{w}}$, for the three different components, alcohol, oil and water, respectively.

In a bulk phase (full of either the oil-rich or water-rich liquid phase or the vapour phase), the grand potential of the system $\Omega_{\textrm{bulk}} = -P V$.
For a given set of parameters that gives rise to phase coexistence, $\Omega_{\textrm{bulk}}$ is the same no matter which phase is chosen. The interfacial tension (excess free energy) between two chosen phases is then the difference between the grand potential for a system of the same volume also having an interface and one without, divided by the area $A$ of the interface, \cite{rowlinson1982molecular, hansen2013theory}
\begin{align}
    \gamma = \frac{\Omega - \Omega_{\textrm{bulk}}}{A}.
    \label{eq:tension_definition}
\end{align}

For a two-phase system (such as liquid water coexisting with its vapour phase) on a solid substrate, three interfacial tensions (solid-liquid, solid-vapour and liquid-vapour) may be calculated and the well-known Young equation relates these to a contact angle formed between the liquid-vapour interface and the solid in the case of a droplet on the substrate \cite{rowlinson1982molecular}.

\begin{figure}[t]
    \centering
    \includegraphics[width=0.75\linewidth]{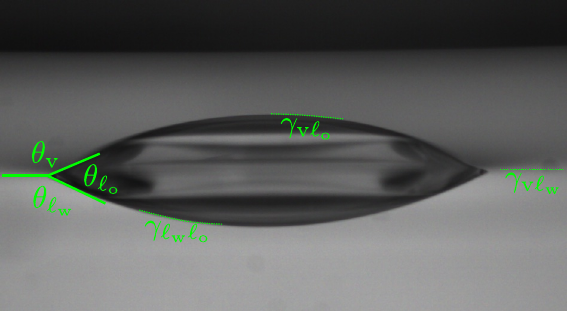}
    \caption{Photograph of a droplet of trans-anethole on water, used to show the surface tensions and Neumann angles of our three-phase system.}
    \label{fig:neumann}
\end{figure}

For our liquid-liquid-vapour system we similarly can calculate three interfacial tensions, i.e.~those between water-rich liquid and oil-rich liquid ($\gamma_{\ell_{\rm w}\ell_{\rm o}}$), between oil-rich liquid and vapour ($\gamma_{{\rm v}\ell_{\rm o}}$), and between water-rich liquid and vapour ($\gamma_{{\rm v}\ell_{\rm w}}$).
Note our use here of the subscripts $\ell_{\rm o}$ and $\ell_{\rm w}$ to denote the oil-rich liquid and water-rich liquid, respectively, rather then just writing `o' or `w', to remind ourselves these are not the pure liquids.
The generalised Young equation becomes three equations formed on Neumann's triangle, see Fig.~\ref{fig:neumann}, and hence the angles through each phase are termed Neumann angles. For our system the equations are \cite{rowlinson1982molecular}
\begin{align}
\label{eq:neumannang1}
\gamma_{{\rm v}\ell_{\rm w}} + \gamma_{\ell_{\rm w}\ell_{\rm o}}\cos\theta_{\ell_{\rm w}} + \gamma_{{\rm v}\ell_{\rm o}}\cos\theta_{\rm v} &= 0,\\
\gamma_{{\rm v}\ell_{\rm o}}\cos\theta_{\ell_{\rm w}} + \gamma_{\ell_{\rm w}\ell_{\rm o}} + \gamma_{{\rm v}\ell_{\rm o}}\cos\theta_{\ell_{\rm o}} &= 0,\\
\gamma_{{\rm v}\ell_{\rm o}}\cos\theta_{\rm v} + \gamma_{\ell_{\rm w}\ell_{\rm o}}\cos\theta_{\ell_{\rm o}} + \gamma_{{\rm v}\ell_{\rm o}} &= 0,
\label{eq:neumannang3}
\end{align}
where $\theta_{\ell_{\rm o}}$, $\theta_{\ell_{\rm w}}$ and $\theta_{\rm v}$ are the Neumann angles through the oil-rich liquid, the water-rich liquid and the vapour phase, respectively.

\section{Lattice DFT Model}
\label{sec:lattice}

In our previous work \cite{archer2024experimental}, we introduced a lattice DFT model for the ouzo system, where the molecules are treated as occupying sites on a discrete lattice. This built on past work using lattice models for various bulk and inhomogeneous liquid systems \cite{kierlik2001capillary, woo2001mean, gouyet2003description, woywod2003phase, robbins2011modelling, monson2012understanding,
schneider2014filling, hughes2014introduction, hughes2015liquid, chacko2015two, chalmers2017dynamical, kikkinides2022connecting, archer2023stability, areshi2024binding}.
In Ref.~\onlinecite{archer2024experimental} the vapour phase is neglected, enabling a thorough investigation of the thermodynamics of the ouzo system with only relatively basic physical properties included.
Whilst we use the same DFT here, we consider also the vapour phase and mostly assume that the liquid phases are in phase coexistence with the vapour (i.e.\ at the triple point). This allows us to study oil droplets at the interface between water (or ouzo) and its vapour and other such intrinsically three-phase phenomena.

\begin{figure}[t]
    \centering
    \includegraphics[width=0.7\linewidth]{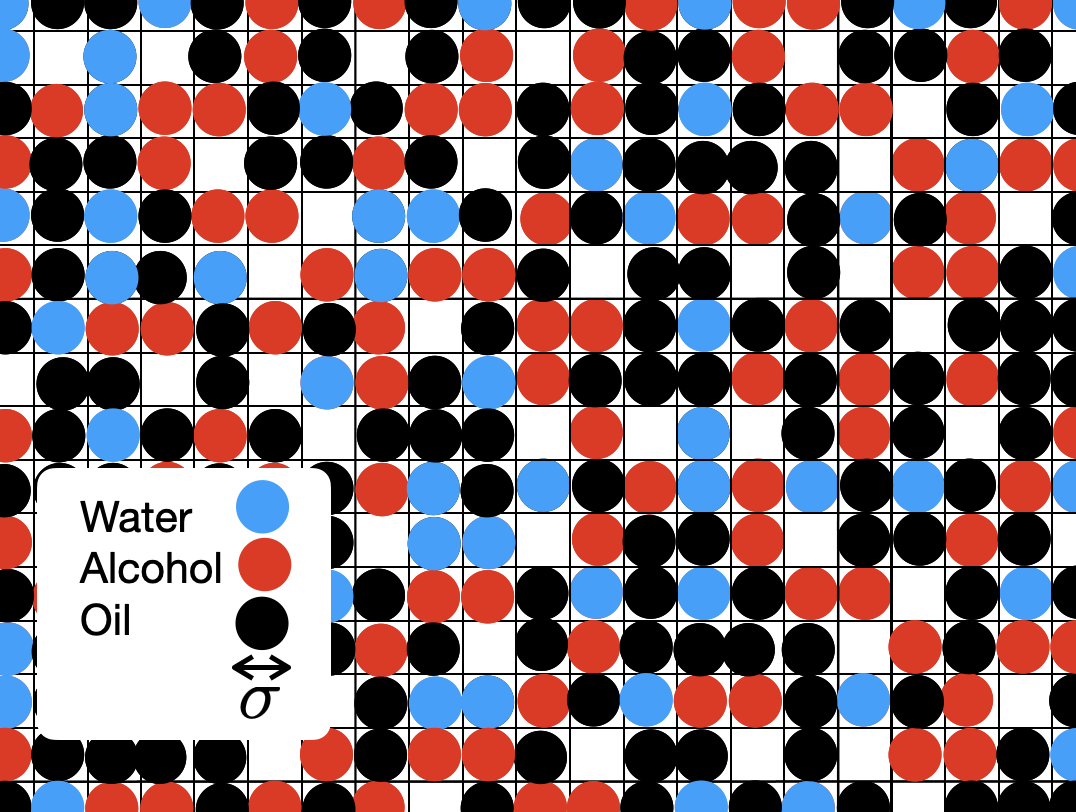}
    \caption{Sketch of the lattice occupied by the various different species used to develop our DFT model, noting that the underlying lattice used to develop the model is actually three-dimensional, rather than the two-dimensional picture shown here.}
    \label{fig:lattice}
\end{figure}

To recap, our system is discretised into a lattice with lattice spacing $\sigma$, roughly defined as the average diameter of an occupying molecule. Whilst alcohol, oil and water have different sizes in reality, we nevertheless saw good agreement with experiments in our previous work. A schematic of the lattice is shown in Fig.~\ref{fig:lattice}. For simplicity, we henceforth set our unit of length $\sigma=1$.

 Full details of the model are given in Ref.~\onlinecite{archer2024experimental}, so here we repeat only aspects and equations that are necessary to explain our new results.
 The principal concept behind the model is to write down an approximation for the Helmholtz free energy that may be minimised to find the distribution in space of the equilibrium ensemble-averaged densities, for a given set of system parameters.
 The ensemble-averaged densities are denoted by $n_{\ii}^{\rm a}$, $n_\ii^{\rm o}$, and $n_\ii^{\rm w}$, where the index $\ii$ denotes the location in space of each lattice site. In three-dimensions (3D) this is $\ii =(i,j,k)$, where $i$, $j$ and $k$ are integers.

The Helmholtz free energy is approximated as \cite{archer2024experimental}
\begin{align}
F = & k_B T \sum_\ii \Big[
        n^{\rm w}_\ii \ln n^{\rm w}_\ii
    + n^{\rm a}_\ii \ln n^{\rm a}_\ii
    + n^{\rm o}_\ii \ln n^{\rm o}_\ii
    \nonumber \\
    & \qquad \qquad + (1 - n^{\rm w}_\ii - n^{\rm a}_\ii-n^{\rm o}_\ii) \ln (1 - n^{\rm w}_\ii - n^{\rm a}_\ii-n^{\rm o}_\ii)\Big]
  \nonumber\\
  &- \sum_{\ii,\j}\Big(
\frac{1}{2} \e_{\ii\j}^{\rm ww} n_{\ii}^{\rm w} n_{\j}^{\rm w}
+\frac{1}{2} \e_{\ii\j}^{\rm aa} n_{\ii}^{\rm a} n_{\j}^{\rm a}
+\frac{1}{2} \e_{\ii\j}^{\rm oo} n_{\ii}^{\rm o} n_{\j}^{\rm o} \nonumber\\
&\qquad \qquad
+ \e_{\ii\j}^{\rm wa} n_\ii^{\rm w} n_\j^{\rm a}
+ \e_{\ii\j}^{\rm wo} n_\ii^{\rm w} n_\j^{\rm o}
+ \e_{\ii\j}^{\rm ao} n_\ii^{\rm a} n_\j^{\rm o}
\Big),
\label{eq:helmholtz}
\end{align}
where the six tensors $\e^{\textrm{pq}}_{\ii,\j}$ (with $\{\textrm{p},\textrm{q}\}$ being $\{\rm a,o,w\}$), define the pair interactions between different lattice sites\cite{chalmers2017modelling, archer2024experimental}. Here, $k_B$ is Boltzmann's constant (and $T$ is the temperature, already noted), usually combined as $\beta=(k_B T)^{-1}$. To determine coexistence values for the various phases and compile a phase diagram, we first consider the bulk fluid phase behaviour where the densities are uniform, i.e.~$n_\ii^{\rm w}=n^{\rm w}$, $n_\ii^{\rm a}=n^{\rm a}$ and $n_\ii^{\rm o}=n^{\rm o}$ are constants for all $\ii$. In this case, the Helmholtz free energy per unit volume $f = F/V$ becomes
\begin{align}
\label{eq:f_bulk}
f = & k_BT\big[n^{\rm w} \ln{n^{\rm w}}+n^{\rm a} \ln{n^{\rm a}}+n^{\rm o} \ln{n^{\rm o}}\nonumber\\
& \qquad + (1 - n^{\rm w} - n^{\rm a} - n^{\rm o}) \ln(1 - n^{\rm w} - n^{\rm a} - n^{\rm o})\big]
\nonumber
\\ 
& - \frac{1}{2}\s^{\rm ww}(n^{\rm w})^2
- \frac{1}{2}\s^{\rm aa}(n^{\rm a})^2
- \frac{1}{2}\s^{\rm oo}(n^{\rm o})^2\nonumber\\
&- \s^{\rm wa} n^{\rm w} n^{\rm a}
- \s^{\rm wo} n^{\rm w} n^{\rm o}
- \s^{\rm ao} n^{\rm a} n^{\rm o},
\end{align}
where $\s^{\rm pq}=\sum_{\j}\e_{\ii\j}^{\rm pq}$ is the integrated strength. As before, we follow Refs.~\onlinecite{chalmers2017modelling, chalmers2017dynamical} by setting $\varepsilon_{\mathbf{ij}}^{pq} = \epsilon_{pq} c_{\mathbf{ij}}$, where
\begin{equation}
\label{eq:c_ij_liquids}
c_{\mathbf{ij}}  = 
  \begin{cases} 
   1 & \text{if }\mathbf{j}\in {NN \mathbf{i}}, \\
    \frac{3}{10}    & \text{if }\mathbf{j}\in {NNN \mathbf{i}}, \\
    \frac{1}{20}    & \text{if }\mathbf{j}\in {NNNN \mathbf{i}}, \\
        0   & \text{otherwise},
         \end{cases}
\end{equation}
and $NN\mathbf{i}$, $NNN\mathbf{i}$ and $NNNN\mathbf{i}$ denote the nearest neighbours of $\mathbf{i}$ (on a cubic lattice in 3D there are 6), next nearest neighbours of $\mathbf{i}$ (in 3D there are 12) and next-next nearest neighbours of $\mathbf{i}$ (in 3D there are 8), respectively. The six parameters $\epsilon^{\rm pq}$, where $\{\rm p,\rm q\}\in \{\rm a,o,w\}$, determine the overall strength of the pair potentials. Thus, the values of these potential strength parameters are simply related to the integrated strengths of the pair potentials, via $\s^{\rm pq}=10\epsilon^{\rm pq}$. In Ref.~\onlinecite{archer2024experimental} we found good agreement with the experiments using the following values
\begin{align}
&\beta\epsilon^{\rm ww}=0.96,\hspace{0.5cm} &\beta\epsilon^{\rm wa}=0.84, \nonumber\\
&\beta\epsilon^{\rm aa}=0.78,\hspace{0.5cm}
&\beta\epsilon^{\rm ao}=0.63, \nonumber \\
&\beta\epsilon^{\rm oo}=0.78,\hspace{0.5cm}&\beta\epsilon^{\rm wo}=0.30,
\label{eq:epsilons}
\end{align}
and thus continue using these values here.
Note that these values are not necessarily the best for all state points and mixtures---in Sec.~\ref{sec:eth-wat} below such an example case is shown for the ethanol-water system only. We leave as future work the possibility to optimise the $\epsilon^{\rm pq}$ values (or add other three-body terms) in order to better match the full gamut of behaviours of oil-alcohol-water systems.

Here, when we consider inhomogeneous systems, we simplify by assuming the density distributions are invariant in one of the Cartesian directions, allowing us to perform our calculations in two-dimensions (2D). Hence, we effectively project from 3D to 2D, so that when we e.g.~calculate density profiles for a 2D circular droplet, this corresponds in 3D to the cross section through an infinitely long cylindrical droplet.
Applying this projection on (\ref{eq:c_ij_liquids}), yields the effective 2D potential 
\begin{equation}
\label{eq:c_ij_liquids2D}
c_{\mathbf{ij}}^{\rm 2D}  = 
  \begin{cases} 
   2 & \text{if } \mathbf{j}=\mathbf{i}, \\
   \frac{8}{5}  & \text{if }\mathbf{j}\in {NN \mathbf{i}}, \\
    \frac{2}{5}    & \text{if }\mathbf{j}\in {NNN \mathbf{i}}, \\
        0   & \text{otherwise},
         \end{cases}
\end{equation}
which is used in all our calculations, such as the droplet simulations in Fig.~\ref{fig:initialcomp}(b) and also in later results, as well as for the effectively 1D calculations for planar interfaces that we also present below. For the latter, we calculate in 2D, setting one of the dimensions to be rather narrow, instead of algebraically summing over the direction parallel to the interface (although this would be possible).

All thermodynamic quantities may be obtained from the free energy. Specifically, the three chemical potentials and the pressure are obtained as \cite{rowlinson1982molecular, hansen2013theory}
\begin{equation}
\mu^{\textrm{p}} = \frac{\partial f}{\partial n^{\textrm{p}}},
\label{eq:mu}
\end{equation}
and
\begin{equation}
P = -f+\mu^{\rm w}n^{\rm w}+\mu^{\rm a}n^{\rm a}+\mu^{\rm o}n^{\rm a}
\label{eq:P}.
\end{equation}

\subsection*{Triple-point `binodal' calculations}

We are now in a position to calculate the phase diagram and associated variables for a bulk system. For two (or more) phases to coexist in static equilibrium, there must be equality in the two (or more) phases of quantities that would give rise to motion or a thermodynamic driving force.
Hence, the temperature $T$, pressure $P$ and chemical potentials $\mu^{\rm p}$ of all species $\textrm{p}$ must be equal in the two (or more) coexisting phases.
Temperature is automatically equal in our model, being a given input parameter.
In our previous study \cite{archer2024experimental}, we concentrated on the coexistence between the two liquid phases, and hence had 4 equations (for $P$ and the three $\mu$'s) for the 6 unknowns (the densities of the 3 components in each of the two liquid phases).
For a consistent mathematical system, we thus had to fix the values of two of the chemical potentials and chose $\mu^{\rm o}$ for the entire phase diagram, and varied the value of $\mu^{\rm a}$ to map out the binodal---the coexistence curve of densities, noting that the phase diagrams were qualitatively very similar over a reasonable range of choices of $\mu^{\rm o}$.
Indeed, we went further in our calculations in Ref.~\onlinecite{archer2024experimental}, restricting the system to be incompressible, such that one density in each phase would be determined from the other two and thus having a consistent mathematical system without fixing $\mu^{\rm o}$, and finding once again a qualitatively identical phase diagram.

To additionally investigate the vapour phase, we cannot of course assume that our system is incompressible. We must also recognise the three possible phases, which as above we denote as $\ell_{\rm o}$, $\ell_{\rm w}$ and v for the oil-rich liquid, the water-rich liquid and the vapour phase, respectively.
Hence, the necessary equations for coexistence (coexistence between all three phases being a triple point) are, again, automatically equal temperature, equal pressure in all three phases (2 equations) and equal chemical potentials (6 equations), with 9 unknowns (the three densities in each of the three distinct phases).
Thus, we need to only fix one of the chemical potentials---which is the one that allows us to trace out the triple-point `binodal'.
To put this another way, applying the Gibbs phase rule, with three components and three phases, one should expect just one degree of freedom in an isothermal system to trace out the triple-point line.
To summarise, our 9 conditions (the 8 above equations with one split by fixing $\mu^{\rm a}$) are
\begin{align}
P_{\ell_{\rm o}} &= P_{\ell_{\rm w}}, &
P_{\ell_{\rm o}} &= P_{\rm v},&
\nonumber\\
\mu_{\ell_{\rm o}}^{\rm o}&=\mu_{\ell_{\rm w}}^{\rm o}, &
\mu_{\ell_{\rm o}}^{\rm o}&=\mu_{\rm v}^{\rm o},&
\nonumber\\
\mu_{\ell_{\rm o}}^{\rm w}&=\mu_{\ell_{\rm w}}^{\rm w}, &
\mu_{\ell_{\rm o}}^{\rm w}&=\mu_{\rm v}^{\rm w},&
\nonumber\\
\mu_{\ell_{\rm o}}^{\rm a}&=\mu_{\rm in}^{\rm a}, &
\mu_{\ell_{\rm w}}^{\rm a}&=\mu_{\rm in}^{\rm a}, & 
\mu_{\rm v}^{\rm a}&=\mu_{\rm in}^{\rm a},
\label{eq:coex_conditions3}
\end{align}
where $\mu_{\rm in}^{\rm a}$ denotes the specified input value of the alcohol chemical potential. These nine equations, taken together with Eqs.~\eqref{eq:f_bulk}, \eqref{eq:mu} and \eqref{eq:P}, are then numerically solved for the nine densities $n^{\rm a}_{\ell_{\rm o}},n^{\rm o}_{\ell_{\rm o}},n^{\rm w}_{\ell_{\rm o}},n^{\rm a}_{\ell_{\rm w}},n^{\rm o}_{\ell_{\rm w}},n^{\rm w}_{\ell_{\rm w}},n^{\rm a}_{\rm v},n^{\rm o}_{\rm v},n^{\rm w}_{\rm v}$.  Given these coexisting densities in the three phases, we can then substitute back into e.g.~Eq.~\eqref{eq:mu} to give the corresponding values of the chemical potentials. Note that, whilst very similar, there is not exact agreement between the coexistence curves for the liquid phases ($\ell_{\rm o}$ and $\ell_{\rm w}$) calculated here and in Ref.~\onlinecite{archer2024experimental}. This is because the latter were obtained for a fixed $\mu^{\rm o}$, whereas here $\mu^{\rm o}$ naturally varies along the coexistence curve (a line of triple points).

To solve our system of nonlinear equations, we use Matlab's in-built \emph{fsolve} function, using both step and function tolerances of $10^{-14}$. We are able to make a sensible initial guess for the nine densities for large negative $\mu_{\rm in}^{\rm a}$, taken as $\beta\mu_{\rm in}^{\rm a}=-10$. As in Ref.~\onlinecite{archer2024experimental}, six of the densities are estimated by observing this is close to being a pure oil-water system and the additional three densities are all relatively easily guessed through being small in the vapour phase. We can then iteratively solve with slightly increased $\mu_{\rm in}^{\rm a}$ in each successive iteration, until there are no longer solutions, which occurs at the critical point, where the two liquids become indistinguishable.

While a comprehensive stability analysis (calculation of spinodal points) for all density combinations becomes arduous and somewhat incomprehensible, we can make some interesting discoveries by considering the stability along the triple-point coexistence lines, and also along tie-lines that connect coexisting points. For this, we require the Hessian determinant, which is given as
\begin{align}
\textrm{det}\left(\textbf{H}_{f\left(n^{\textrm{a}},n^{\textrm{o}},n^{\textrm{w}}\right)}\right) = 
\begin{vmatrix}
  \dfrac{\partial^2 f}{\partial n^{\textrm{a}2}} & \dfrac{\partial^2 f}{\partial n^{\textrm{a}}\,\partial n^{\textrm{o}}} & \dfrac{\partial^2 f}{\partial n^{\textrm{a}}\,\partial n^{\textrm{w}}} \\[2.2ex]
  \dfrac{\partial^2 f}{\partial n^{\textrm{o}}\,\partial n^{\textrm{a}}} & \dfrac{\partial^2 f}{\partial n^{\textrm{o}2}} & \dfrac{\partial^2 f}{\partial n^{\textrm{o}}\,\partial n^{\textrm{w}}} \\[2.2ex]
  \dfrac{\partial^2 f}{\partial n^{\textrm{w}}\,\partial n^{\textrm{a}}} & \dfrac{\partial^2 f}{\partial n^{\textrm{w}}\,\partial n^{\textrm{o}}} & \dfrac{\partial^2 f}{\partial n^{\textrm{w}2}}
\end{vmatrix},
\label{eq:hessian}
\end{align}
and is zero at the change of stability. We are now in a position to show our results.

\section{Results}
\label{sec:4}

\subsection{Coexistence lines}

The equations in the preceding sections enable us to present the phase diagram showing where three-phase coexistence occurs.
Traditionally, ternary phase diagrams are displayed in a triangular format (see Ref.~\onlinecite{archer2024experimental} for details of how these projections are obtained), where the distance from each of the three corners is related to the relative concentration of each of the three species.
Therefore, as expected, it was found \cite{archer2024experimental} that the liquid-liquid binodal curves are clear to see and easy to interpret when displayed in this manner.
\begin{figure}[t]
    \centering
    \includegraphics[width=0.99\linewidth]{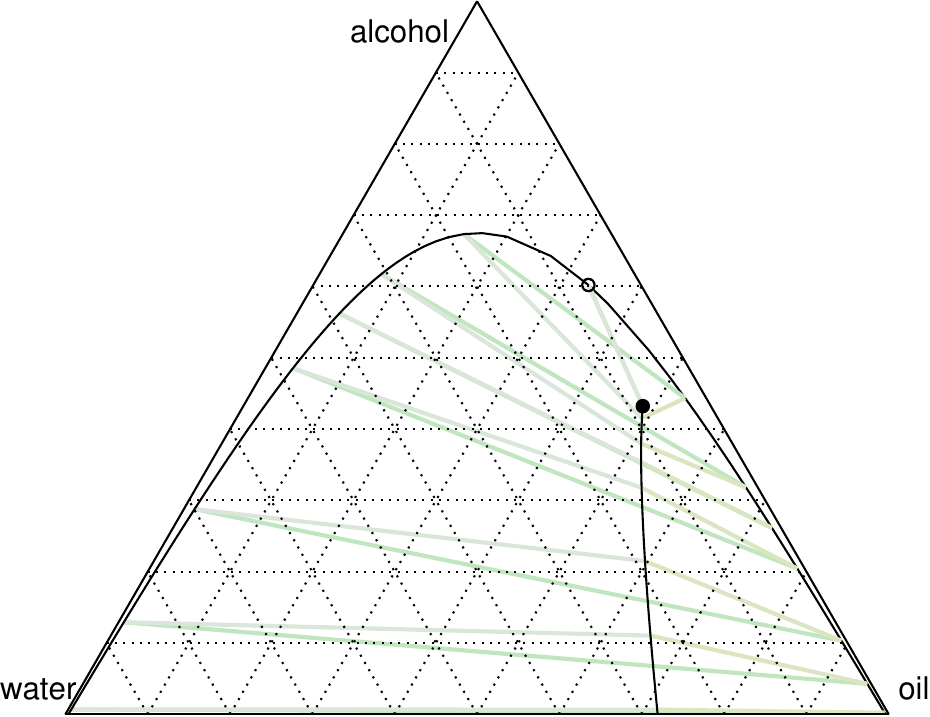}
    \caption{Points of three-phase coexistence displayed in the traditional ternary phase diagram, where the number densities along the coexistence (`binodal') curves are mapped to a corresponding distance from the triangle corners. The three corners of the triangle are locations where 100\% of the written component exist, with the dotted black-line triangular grid being at 10\% intervals in each component. The solid black lines are lines of phase coexistence. The empty circle is the liquid-liquid critical point, and the filled circle is the termination point of the vapour branch (occurring at the same state point as the critical point). The various green lines are tie lines between coexisting state points on the oil-rich liquid, water-rich liquid and vapour branches, respectively, and are described in detail in the main text.}
    \label{fig:TPhaseD}
\end{figure}
However, in the equivalent case here, where we also consider the vapour phase, it is less easy to interpret.
One major drawback of this representation is that densities are converted into ratios.
Consider e.g.\ the central point, corresponding to equal alcohol, oil and water.
This point could be a liquid with each of the number densities equal to say 0.3, or it could correspond to a vapour where each has number density $10^{-3}$.
Indeed, any point in the traditional triangular ternary phase diagram represents an infinite number of scaled triples $(n^{\textrm{a}},n^{\textrm{o}},n^{\textrm{w}})$, where the dimensionless number densities $n^{\rm p}$ are each bounded between 0 and 1 and their sum is also bounded between 0 and 1.
This means that most points in this projection are both stable and unstable state-points according to Eq.~\eqref{eq:hessian}, depending on the specific set of three number densities, and not just their ratios.
We do nevertheless present our results in this way, but alongside other (3D) representations. 
There is some effort required to read these other representations, but we believe the effort is worth it, to fully appreciate the complexity of the phase behaviour.

Firstly, we plot the three-phase coexisting states in the traditional triangular ternary mixture phase diagram in Fig.~\ref{fig:TPhaseD}.
When comparing to the two-phase (liquid-liquid) ternary phase diagram obtained in Ref.~\onlinecite{archer2024experimental}, we now have an additional line of solutions formed by the vapour phase.
Solutions cease when the liquid-liquid critical point is reached (since three phases are needed for three-phase coexistence!), so the vapour branch also terminates at the same state point, but this is not at the same point in the triangular phase diagram.
The critical point is shown with an empty circle, and at the same value of $\mu^{\textrm{a}}$ the termination point on the vapour line is shown with a filled circle. Between the coexisting liquids, we previously drew tie lines for selected points at the same $\mu^{\textrm{a}}$, and do the same here (darkest green).
With the new vapour phase, we now have three tie lines at each $\mu^{\textrm{a}}$ since we need to additionally join points from the water-rich liquid phase to the vapour (lightest green) and from the oil-rich liquid phase to the vapour (middle green).
Due to the projection of solutions onto the triangular ternary phase diagram, this means that we observe triangles except in two cases: one case where the projected lines end up over each other; and the second case at the critical point, where of course instead we can only connect the indistinguishable liquid phase to the vapour phase. We note that the tie-lines are defined as straight lines in density space (i.e.~we connect coexisting points of $(n^{\textrm{a}},n^{\textrm{o}},n^{\textrm{w}})$ linearly, and when projected into the ternary phase diagram they appear to remain as straight lines).

\begin{figure*}[t]
    \centering
    \includegraphics[width=0.7\linewidth]{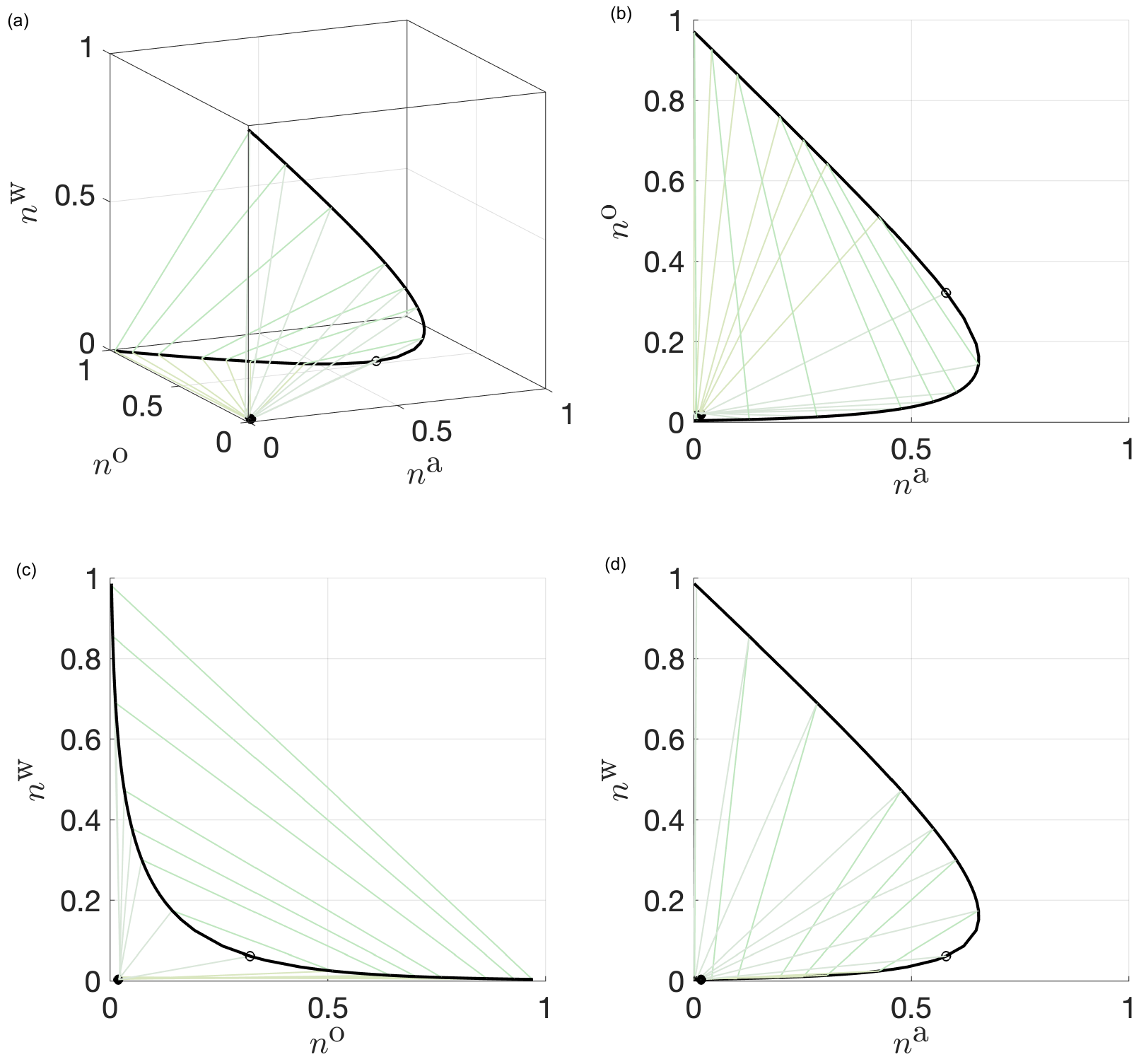}
    \caption{Plots of the three-phase coexistence curves using the actual number densities. (a) shows this in 3D, while (b), (c) and (d) are the same data as (a) but from different viewing angles. Line styles match those in Fig.~\ref{fig:TPhaseD}.}
    \label{fig:3D}
\end{figure*}

We can see the deficiency with this representation when following a tie line between two liquids. Starting from a point on the coexistence line where there is water-rich liquid, then following a selected tie line we expect the water density to slowly decrease being replaced by oil with a small adjustment of alcohol but without much change in total density all the way until we reach the corresponding coexistence line branch where there is oil-rich liquid. This is indeed what happens, but in this visualisation somewhere in the middle we intersect with the line for the coexisting vapour phase.
This is of course purely due to this projection and there is no true intersection in density space. We can more clearly understand this by considering Fig.~\ref{fig:3D}(a), where we plot the coexistence curves as functions of the number densities $(n^{\textrm{a}},n^{\textrm{o}},n^{\textrm{w}})$, in a three-dimensional plot.

\begin{figure*}
    \centering
    (a)\hspace{8.5cm}\vphantom{a} (b)\hspace{7cm}\vphantom{a}\\
    \includegraphics[width=0.49\linewidth]{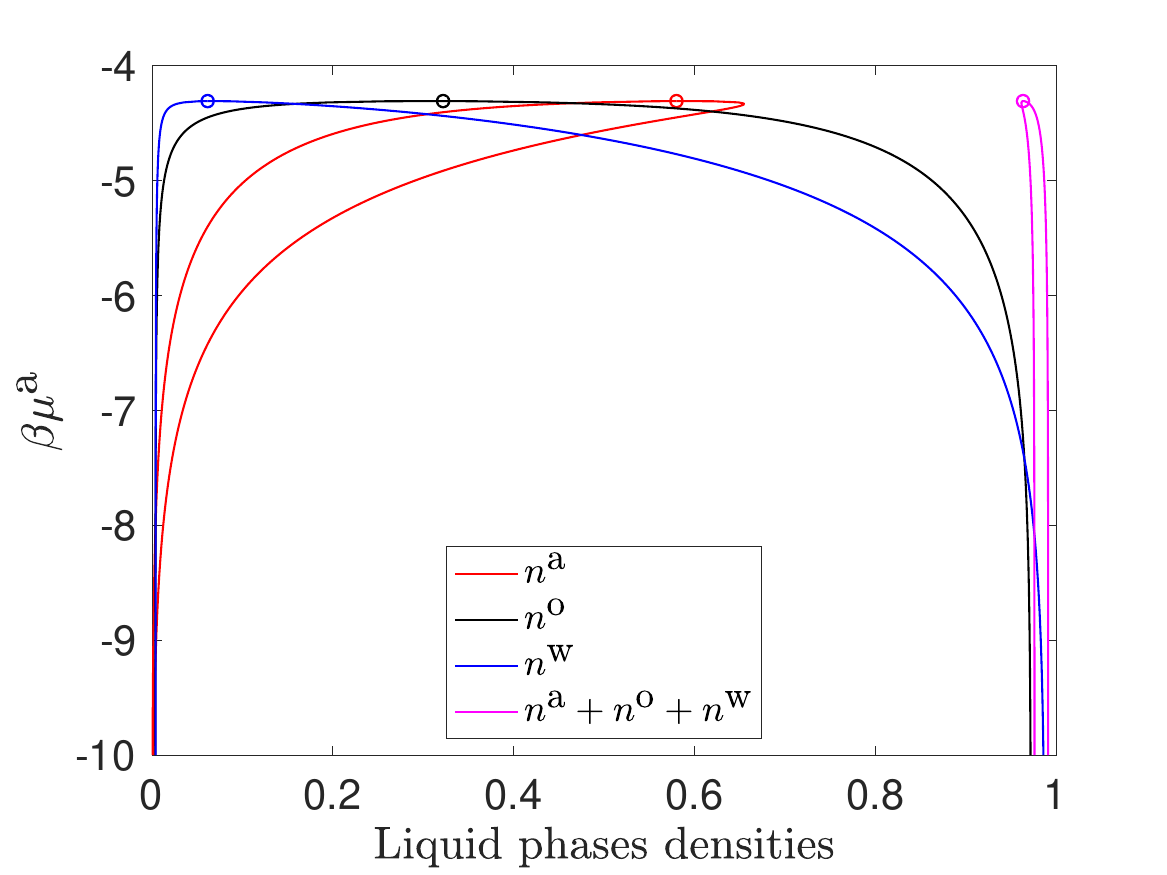}
    \includegraphics[width=0.49\linewidth]{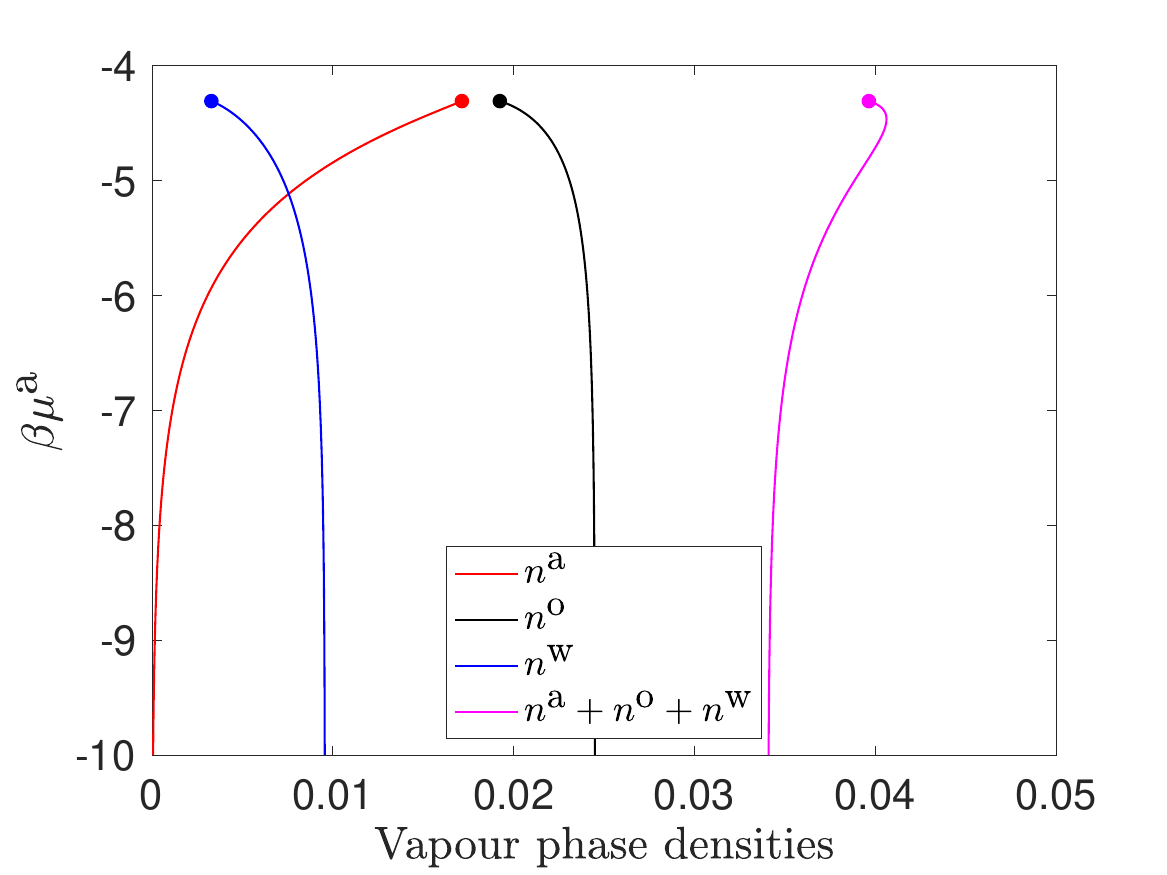}\\\vphantom{a}\\
    (c)\hspace{5.5cm} (d) \hspace{5.5cm} (e) \hspace{4cm}\vphantom{a}\\
    \includegraphics[width=0.32\linewidth]{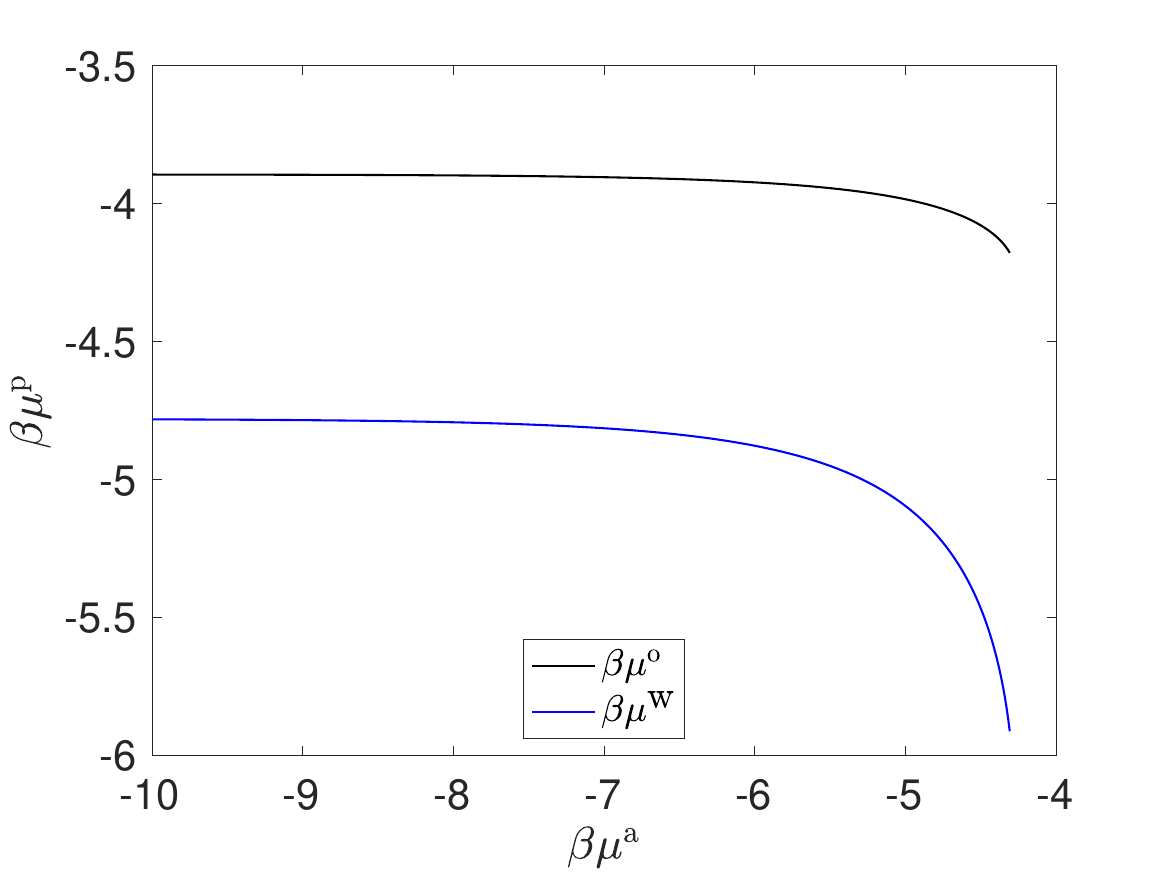}
    \includegraphics[width=0.32\linewidth]{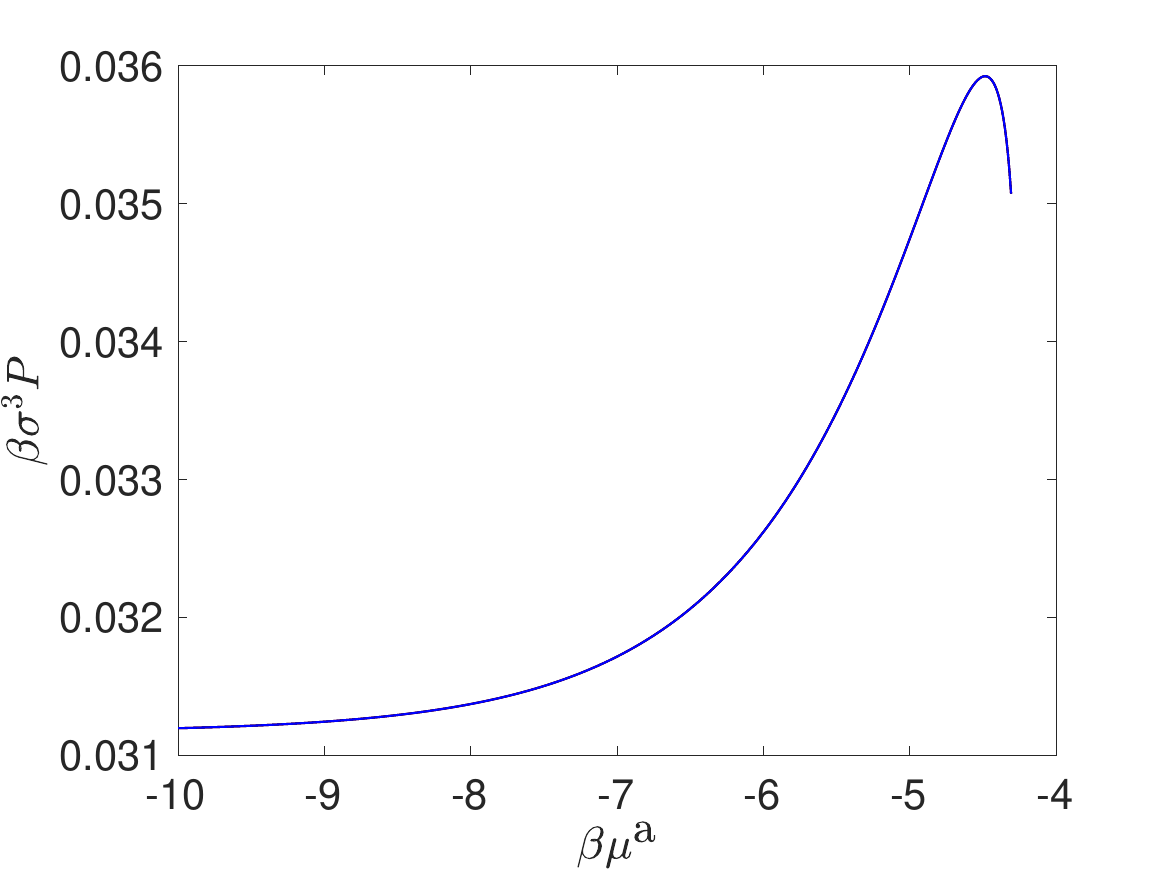}
    \includegraphics[width=0.32\linewidth]{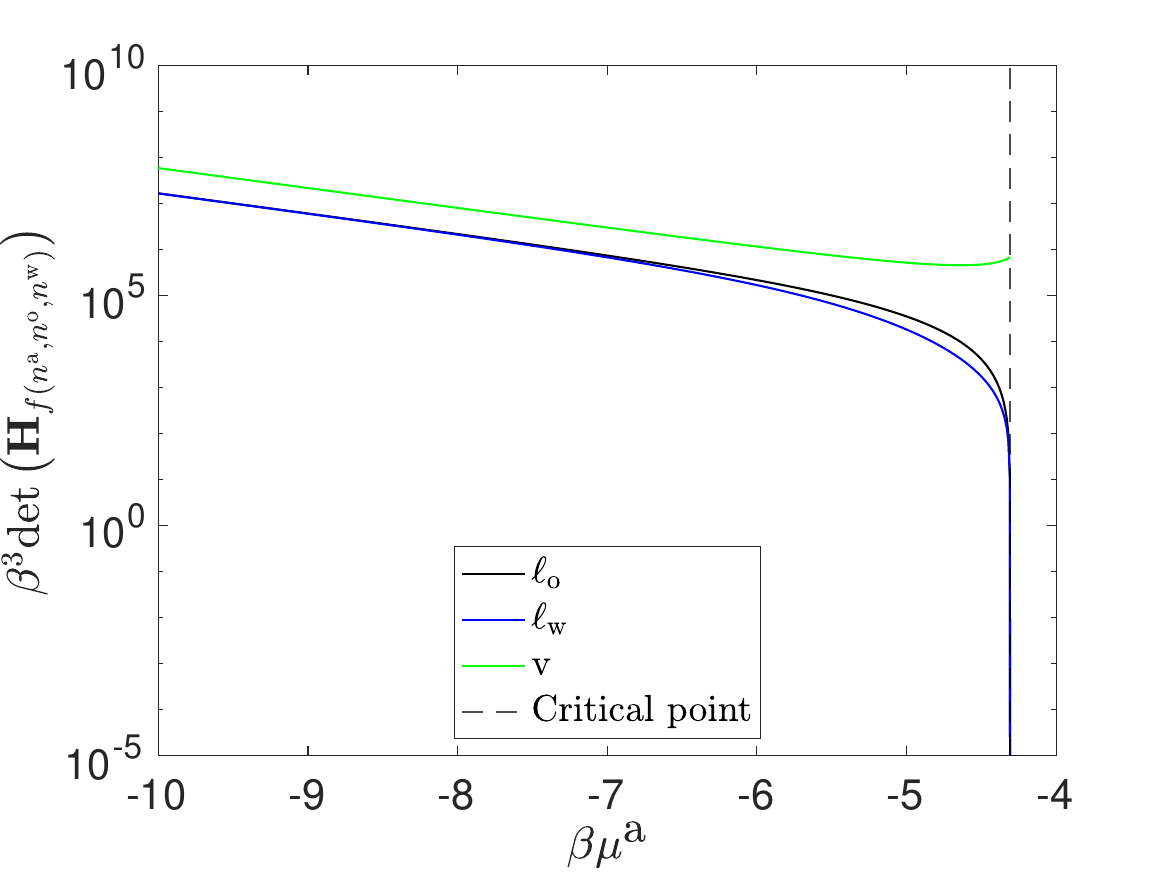}
    \caption{Plots of the densities of each species along the three-phase coexistence for (a) the liquid phases and (b) the vapour phase,  as $\mu^{\textrm{a}}$ is varied to trace out the coexistence curves. We also display other thermodynamic quantities along the three-phase coexistence: In (c) the varying chemical potentials, (d) the pressure, and (e) the Hessian determinant from (\ref{eq:hessian}).}
    \label{fig:rhomu}
\end{figure*}

In Fig.~\ref{fig:3D} we display the tie lines using the same colours as used in Fig.~\ref{fig:TPhaseD}.
Figure~\ref{fig:3D} shows clearly that the tie lines between the coexisting liquid phases go nowhere near the vapour phase, which, due to all the densities being quite small for all state points, is quite a short line near the origin in this representation.
Likewise, the tie lines between the vapour and one of the two liquids are well away from the states corresponding to the other liquid, except at the critical point.
Figures~\ref{fig:3D}(b--d) are exactly the same as Fig.~\ref{fig:3D}(a) but displayed from different viewing angles, since curves in 3D representations can be difficult to properly appreciate at a single angle.

In order to trace out the coexistence line curves, we vary $\mu^{\textrm{a}}$, starting from $\beta\mu^{\textrm{a}}=-10$, where there is almost no alcohol in the system and terminating at the critical point.
Thus, another instructive visualisation is seeing how the densities change with $\mu^{\textrm{a}}$; these are displayed in Fig.~\ref{fig:rhomu}(a) and (b). In Fig.~\ref{fig:rhomu}(c) we also show how the other two chemical potentials correspondingly change along the three phase coexistence line. Note that these are not directly solved for, but are obtained from the density solutions using Eq.~\eqref{eq:mu}. In a similar way, we can also calculate the pressure using Eq.~\eqref{eq:P}. Of course, the result is the same for all three coexisting phases; the resulting curves all lie on top of each other (see Fig.~\ref{fig:rhomu}(d)).

In Fig.~\ref{fig:rhomu}(e) we plot the Hessian determinant in Eq.~\eqref{eq:hessian} calculated along the coexistence curves. This of course goes to zero at the liquid-liquid critical point along the liquid state branches, but not on the vapour phase branch, because this determinant is proportional to the inverse of the compressibility.
This signifies that the vapour phase remains distinguishable from the liquid phase even though the 2 liquid phases are identical.

\subsection{Spinodal lines}

\begin{figure*}
    \centering
    \includegraphics[width=0.75\linewidth]{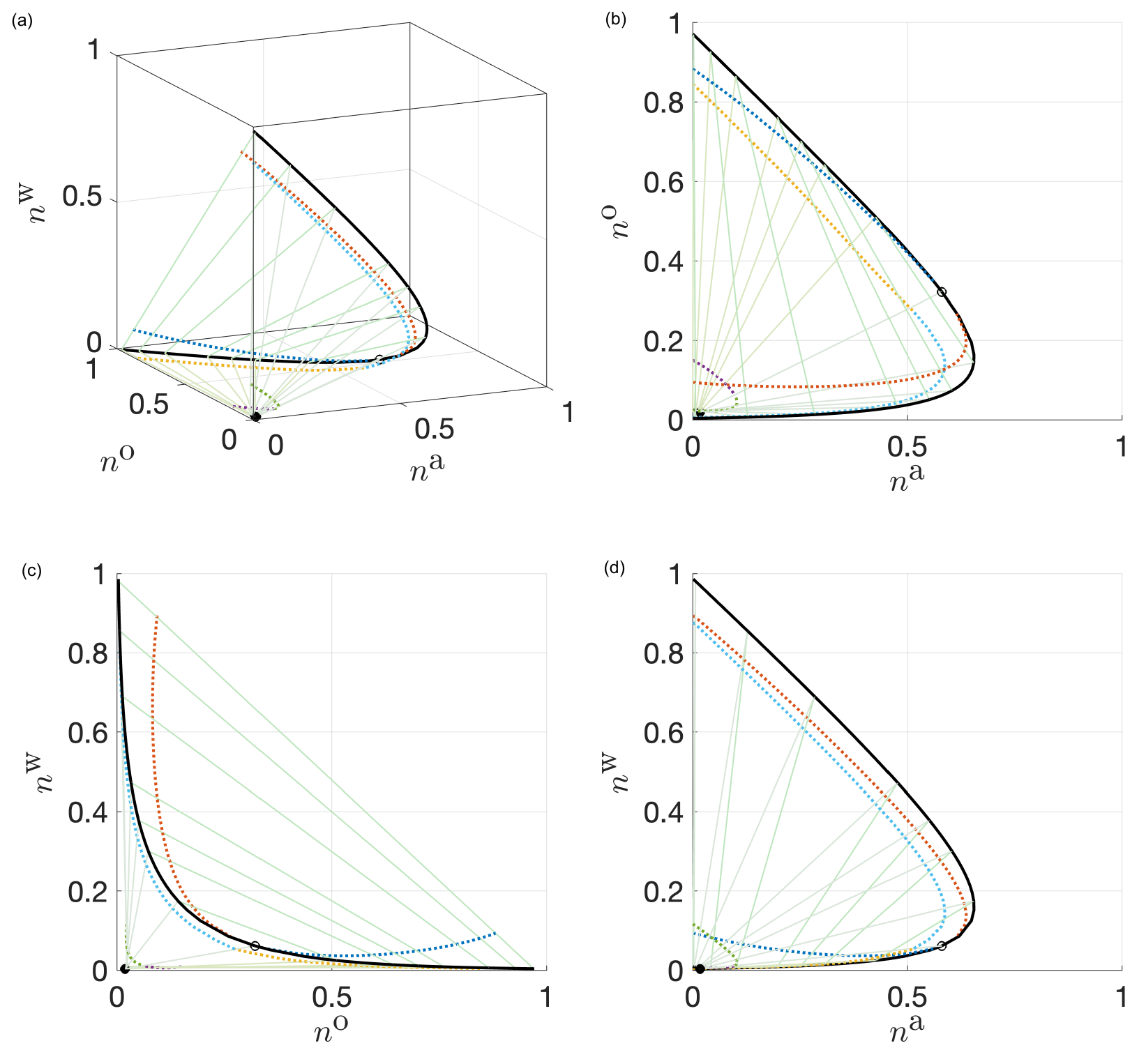}
    \caption{Plots of the coexistence curves (black solid lines) together with `tie-line spinodal' curves using the actual number densities. This plot is the same as Fig.~\ref{fig:3D}, except here we also include the lines of spinodal points calculated along the tie-lines. The red and dark-blue dotted lines are the spinodal points calculated along the liquid-liquid coexistence tie-lines, while the others are spinodals on the respective liquid-vapour coexistence tie-lines, for both the oil-rich-liquid and also the water-rich-liquid phases. (a) shows these in 3D, while (b), (c) and (d) are also the same as (a) but from different viewing angles. }
    \label{fig:3Dspin}
\end{figure*}

\begin{figure}
    \centering
    (a)\hspace{7.5cm}\vphantom{a}\\ \includegraphics[width=0.99\linewidth]{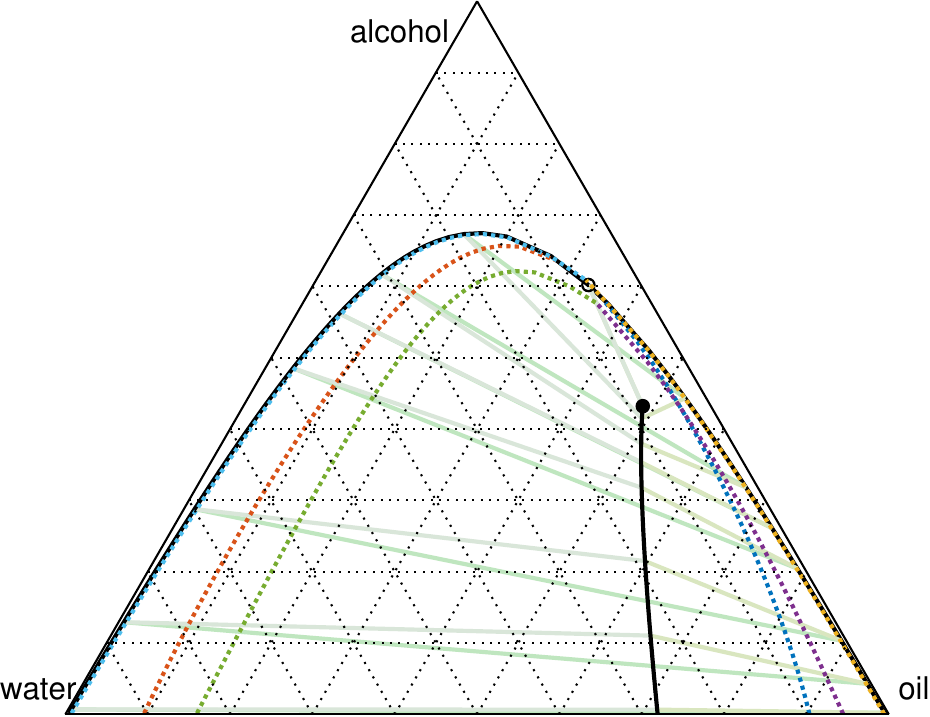}\\
    (b)\hspace{7.5cm}\vphantom{a}\\ \includegraphics[width=0.99\linewidth]{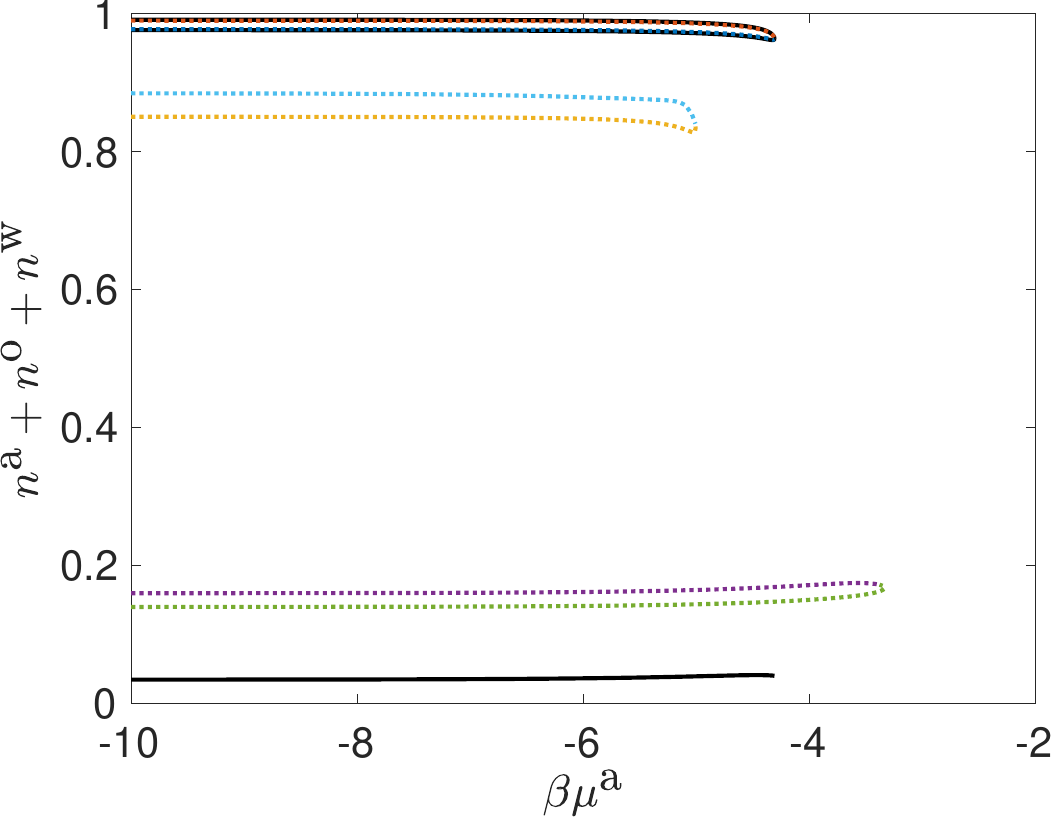}
    \caption{(a) The traditional ternary phase diagram additionally including the `tie-line spinodal' curves, and (b) the total density along these curves. For tie-lines between the two liquid phases, stability changes at the dark-blue and red dashed lines, and these both meet at the critical point. For the oil-rich liquid-vapour lines, stability changes on the yellow and purple dashed lines and the corresponding water-rich liquid-vapour lines are dashed light-blue and green. Unlike liquid-liquid spinodals, the vapour-related ones meet in a different fashion---they don't meet their same-phase counterpart since the vapour phase coexistence curve does not intersect with any other phase curve, but they do meet their opposing counterpart on the tie-line connecting the vapour termination point and the critical point.}
    \label{fig:TPhaseDspin}
\end{figure}

As mentioned in the previous section, it is challenging to comprehensively present all the regions of stability in the phase diagram of our system.
Looking at Fig.~\ref{fig:3D}, we can in principle calculate the stability of every point in $(n^{\textrm{a}},n^{\textrm{o}},n^{\textrm{w}})$ space, and thus the resulting surfaces at the change of stability, i.e.\ the spinodal surfaces.
To make meaningful headway, and given that experimentally we are often interested in mixtures of coexisting phases, we instead restrict our search to the changes of stability along tie-lines.
In other words, we calculate the point along all of the tie-lines where the Hessian determinant \eqref{eq:hessian} equals zero.
We repeat Figs.~\ref{fig:3D} and \ref{fig:TPhaseD} in Figs.~\ref{fig:3Dspin} and \ref{fig:TPhaseDspin}, supplementing the latter with a total density plot that helps deconstruct the flattened projection of the ternary phase diagram.


The complexity of these spinodal surfaces is evident in Fig.~\ref{fig:3Dspin}(a).
It is interesting to see that the spinodals between vapour and each of the liquid phases are quite far apart in density-space, signifying that one of the two changes in stability happens `closer' to the vapour phase (i.e.~with total density much lower) than the other.
Much more strikingly, the projection onto the classical ternary phase diagram displayed in Fig.~\ref{fig:TPhaseDspin}(a), has one of these spinodals lying almost over the top of the coexistence curve---even though they are quite clearly distinct in density-space.
This means that the triangular ternary phase diagram representation could be misleading because it can appear as if essentially everything within the liquid-liquid coexistence line is unstable, even though in reality this isn't the case.
This is purely due to the effect of the projection.
We help unpick this through plotting (in corresponding colours) the total density of coexistence and spinodal curves in Fig.~\ref{fig:TPhaseDspin}(b), where we see the light blue-yellow spinodal is quite separate from the black coexistence line (which happens to have almost identical total density to the dark blue-red spinodal, but this isn't surprising since these are spinodals along the liquid-liquid coexistence tie-lines, and the total density doesn't vary all that much along these curves).
The proximity of one of the spinodals to the liquid-liquid coexistence line is interesting.
A consequence of this is that in this way of displaying the phase diagram, we could have a region very close to this `binodal' where bubbles could cavitate as the liquid phase becomes unstable past the spinodal line. This region does exist, but occurs at a lower total density such that nucleating bubbles could form but where initial densities are closed to those of the vapour phase.
Note that for Fig.~\ref{fig:TPhaseDspin}(b) we have run our full phase-diagram computation to values smaller than $\beta\mu^{\rm a}=-10$ to capture the spinodal curves to this value.

\subsection{Surface tensions}

\begin{figure*}
    \centering
    \includegraphics[width=0.99\linewidth]{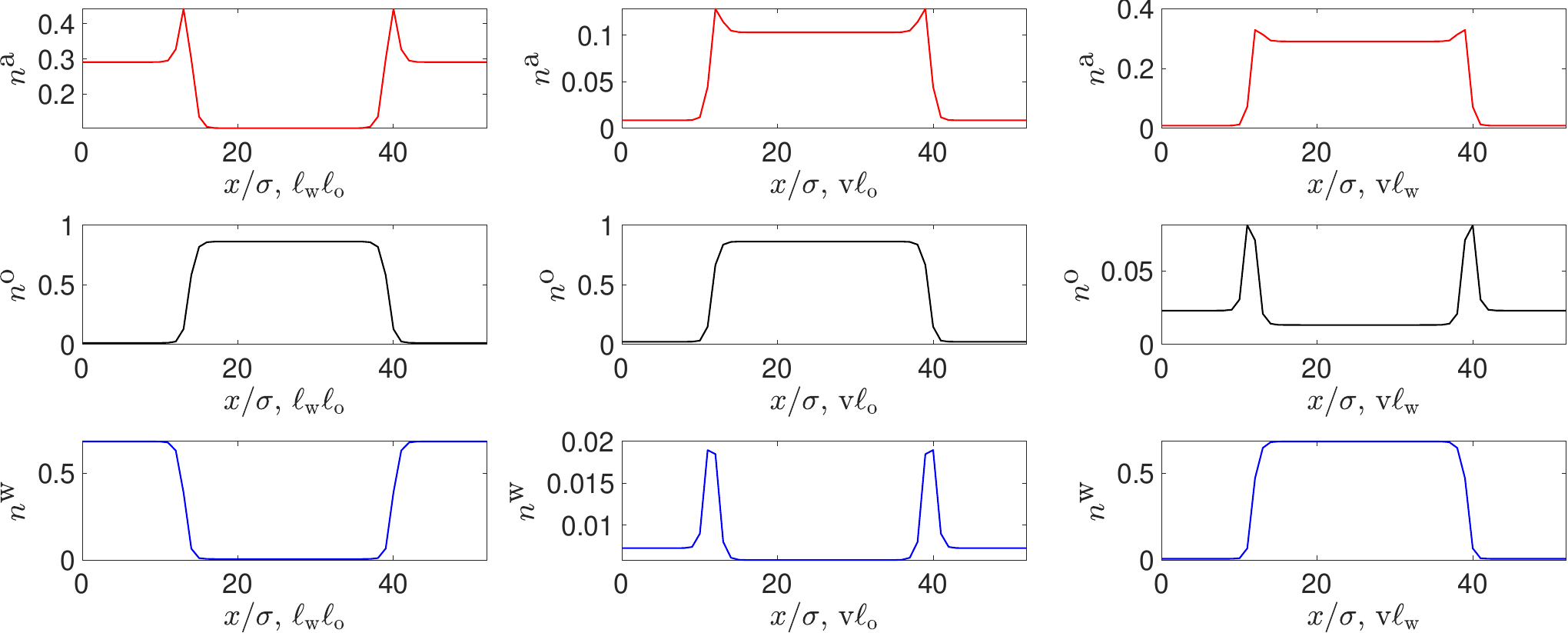}
    \caption{Density profiles for the planar interface between the 3 possible coexisting phases at three-phase coexistence, for the case where $\beta\mu^{\rm a}=-5$ (see also Fig.~\ref{fig:initialcomp}). Note that the vertical scales vary from plot to plot. Moreover, the resolution of these profiles is set by the lattice (with $\sigma=1$), and hence interfaces are not smooth by design of the model rather than by computational parameters. The top row shows the alcohol density profiles, the middle those for the oil and the bottom row the water profiles, corresponding to the interface between the oil-rich-liquid $\ell_{\rm o}$ and the water-rich-liquid $\ell_{\rm w}$ (left), the interface between $\ell_{\rm o}$ and the vapour v (middle) and the interface between $\ell_{\rm w}$ and v (right). Note that in all three case the occurrence of density peaks (surface excess adsorptions) of minority species at the interfaces.}
    \label{fig:profiles_m5}
\end{figure*}

To compute the surface tensions of the various interfaces we follow essentially the same procedure as outlined in Refs.~\onlinecite{hughes2014introduction, archer2024experimental}.
We take the computed triplet of density values and corresponding three chemical potential values for each of a pair of coexisting phases and create an initial condition that corresponds to the interface between these. Actually, our system has two interfaces, due to the use of periodic boundary conditions.
Referring to the two coexisting phases as A and B, we create an effectively one-dimensional initial A-B-A profile (though, we actually use a two-dimensional set-up that is narrow in the direction perpendicular to the interfaces and is doubly-periodic).
Starting from this, we use Picard iteration to minimise the grand potential
\begin{equation}
\Omega=F
-\sum_\ii\mu^{\rm a}n_\ii^{\rm a}
-\sum_\ii\mu^{\rm o}n_\ii^{\rm o}
-\sum_\ii\mu^{\rm w}n_\ii^{\rm w}
,
\label{eq:Omega}
\end{equation}
i.e.~we solve
\begin{equation}
\frac{\delta \Omega}{\delta n_\ii^{p}}=0,
\label{eq:dOmega}
\end{equation}
for $p=\{{\rm a,o,w}\}$, where we find a mixing parameter of $0.001$ and convergence criterion of $\sum_{p \in \{{\rm a,o,w}\}}\sum_\mathbf{i}(n_{\mathbf{i},\mathrm{new}}^\mathrm{p}-n_{\mathbf{i},\mathrm{old}}^\mathrm{p})^2<10^{-14}$ is sufficient, taking (often substantially) fewer than an assigned maximum number of $10^8$ steps.
Once convergence is reached, this yields equilibrium profiles containing two one-dimensional interfaces between the two chosen coexisting phases. Figure~\ref{fig:profiles_m5} displays the results for the case where $\beta\mu^{\rm a}=-5$, chosen as was also the case earlier in Fig.~\ref{fig:initialcomp}, showing the profiles for each of the three species (rows) and for the case where the coexisting fluids are the water-rich liquid and the oil-rich liquid (left column), the oil-rich liquid and the vapour (middle column) and the water-rich liquid and the vapour (right column).

Regions of the density profiles where the $n^\textrm{p}_\ii$ values are constant correspond to the bulk phases that occur at the vertices of the chosen tie-line triangle in the phase diagrams.
The portions where $n^\textrm{p}_\ii$ varies correspond to the interfaces between these bulk phases.
The profiles of course vary between different coexisting state points in the phase diagram (varied by changing $\beta\mu^{\rm a}$), and especially close to the critical point where the liquid-liquid interface substantially broadens. However, the ones displayed in Fig.~\ref{fig:profiles_m5} are to some extent representative and a more obvious feature, as mentioned in the caption, is the enhancement of the minority species at the interface.
A future direction of our research is to understand whether this is related to the stability of oil-rich droplets in emulsified ouzo.

\begin{figure*}
    \centering
    (a) \hspace{8.5cm} (b) \hphantom{AAAAAAAAAAAAAAAAAAAAAAAAAAAAAAAA}
    \\
    \includegraphics[width=0.49\linewidth]{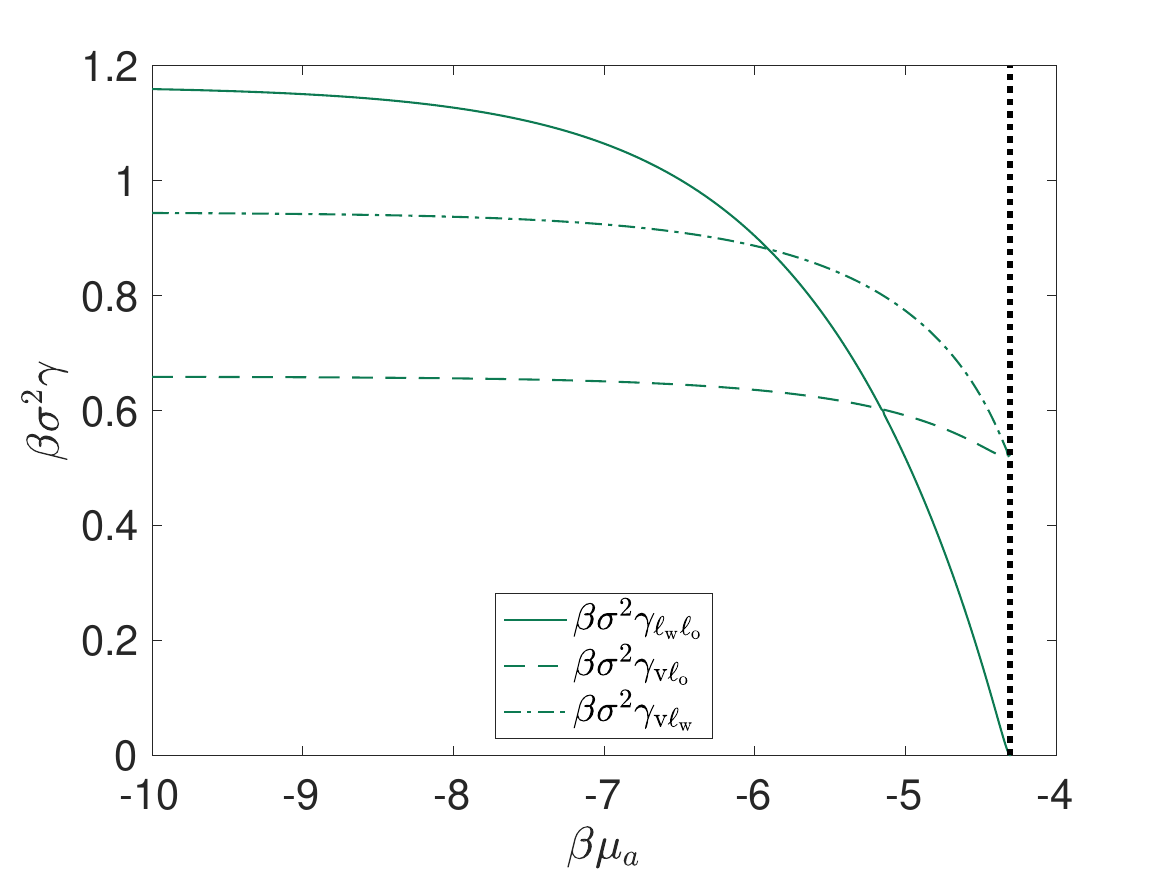} \,
        \includegraphics[width=0.49\linewidth]{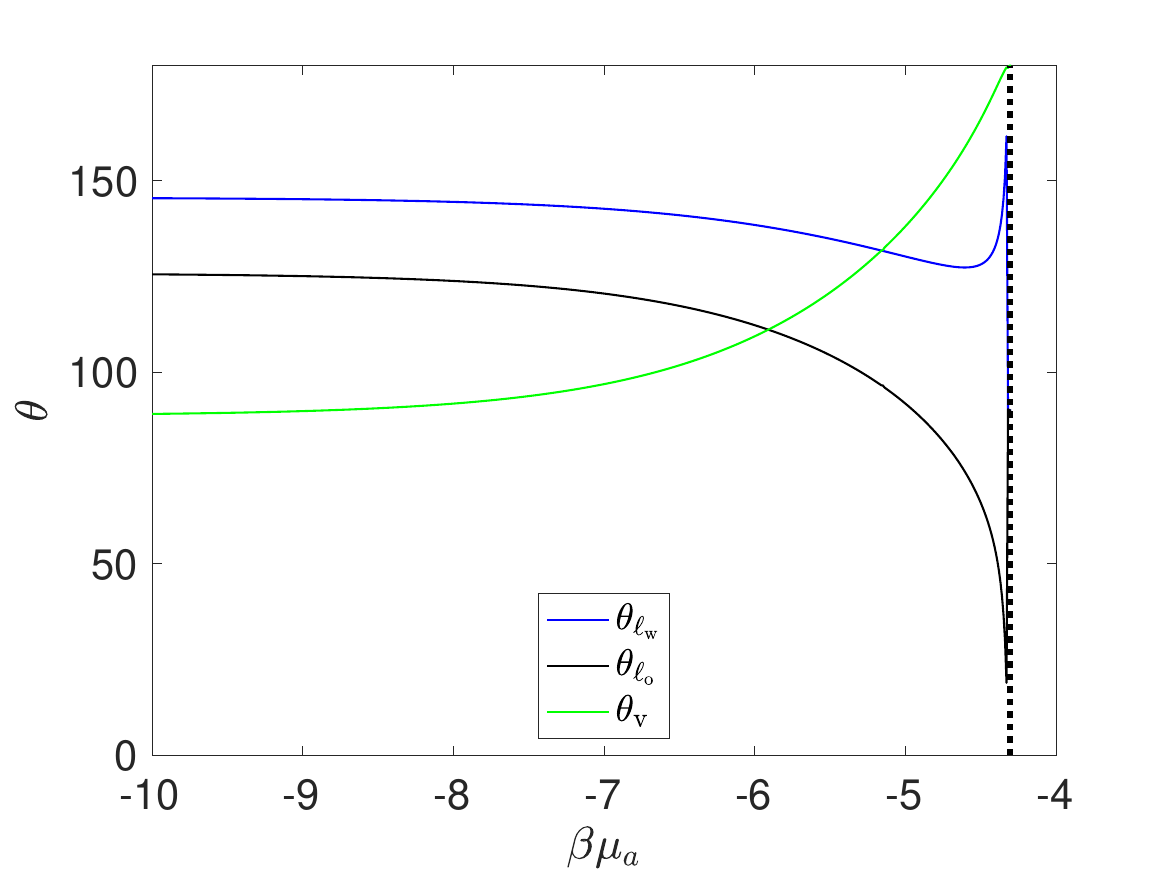}
    \caption{(a) The surface tensions computed from the relevant density profiles (one specific set of profiles shown in Fig.~\ref{fig:profiles_m5}) for the three possible 2-phase combinations up to the liquid-liquid critical point. (b) The three Neumann angles through each phase, computed using (\ref{eq:neumannang1})-(\ref{eq:neumannang3}) with the surface tensions in (a).}
    \label{fig:thetsgams}
\end{figure*}

Once we have the equilibrium density profiles, we can directly compute the interfacial tensions (excess free energies) via Eq.~\eqref{eq:tension_definition} together with Eq.~\eqref{eq:Omega}, and subsequently the Neumann angles using Eqs.~\eqref{eq:neumannang1}-\eqref{eq:neumannang3}. These are shown in Fig.~\ref{fig:thetsgams}(a) and Fig.~\ref{fig:thetsgams}(b), respectively.
As mentioned, as the critical point at $\beta\mu^{\textrm{a}}\approx-4.3087$ between liquid phases is approached by increasing $\beta\mu^{\textrm{a}}$, the interface between the liquid phases broadens and ultimately the phases become indistinguishable, with the surface tension between the two liquid phases $\gamma_{\ell_{\rm w}\ell_{\rm o}}\to0$.
In the vicinity of the critical point, all of the surface tensions and the resulting Neumann angles (which tend to either zero or 180$^\circ$) become more challenging to compute, requiring larger system sizes to be accurate.

\begin{figure}[t]
    \centering
    \includegraphics[width=0.99\linewidth]{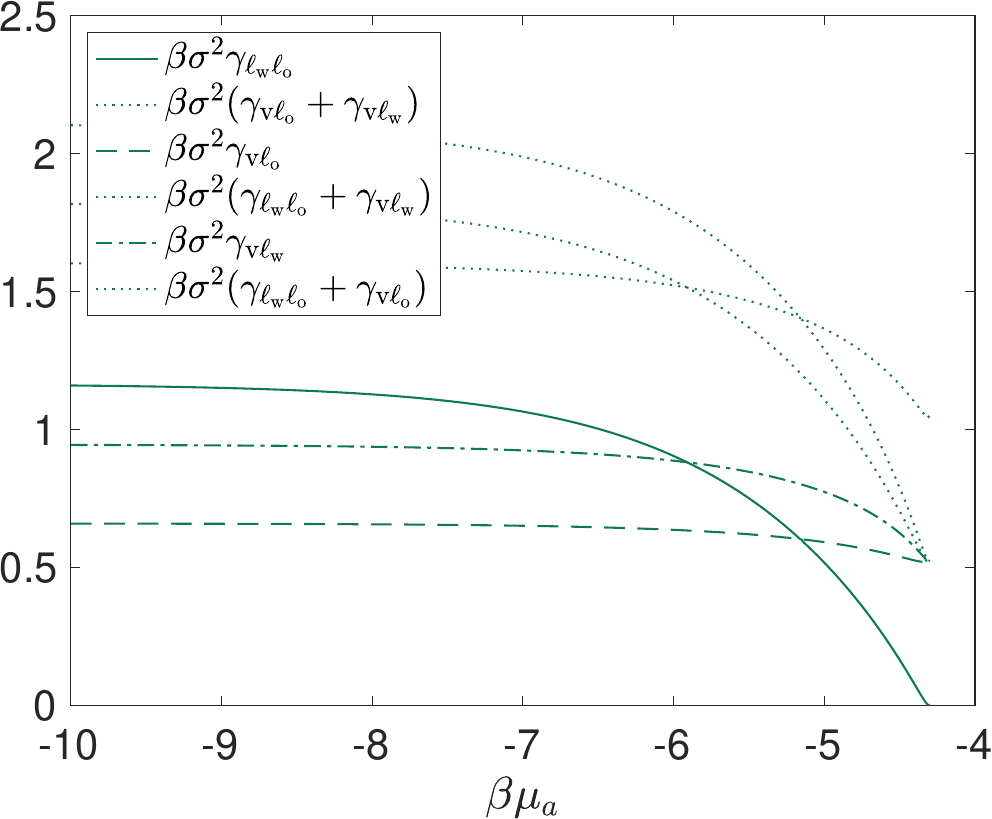}
    \caption{Comparison of the three surface tensions and sums of pairs of these [c.f.~Eqs.~\eqref{eq:inequalities}], to show it is never energetically favourable to have three phases together compared to two.}
    \label{fig:ifwet}
\end{figure}

One interesting corollary from our surface tension results is that we can predict whether it is ever energetically favourable for a third phase to wet the interface between the other two.
For example, if the oil-rich liquid and the vapour phases are placed together, one might wonder if the system would evolve towards having a `droplet' of the water-rich phase condense between them?
Given the density enhancement at interfaces of some components, this is perhaps a natural question to ask.
However, we can confirm that it is never preferable to have a third bulk phase in the middle of the other two, through confirming that
\begin{align}
\gamma_{{\ell}_{\mathrm{w}}{\ell}_{\mathrm{o}}}
<
\gamma_{\mathrm{v}{\ell}_{\mathrm{o}}}+\gamma_{\mathrm{v}{\ell}_{\mathrm{w}}}, \nonumber \\
\gamma_{\mathrm{v}{\ell}_{\mathrm{o}}}<
\gamma_{{\ell}_{\mathrm{w}}{\ell}_{\mathrm{o}}}+\gamma_{\mathrm{v}{\ell}_{\mathrm{w}}},\nonumber \\
\gamma_{\mathrm{v}{\ell}_{\mathrm{w}}}<
\gamma_{{\ell}_{\mathrm{w}}{\ell}_{\mathrm{o}}}+\gamma_{\mathrm{v}{\ell}_{\mathrm{o}}},
\label{eq:inequalities}
\end{align}
which is also illustrated in Fig.~\ref{fig:ifwet}.
Thus, the observed enhancements of the minority species at the interfaces is not a sign of incipient wetting, and we believe is due to being at the triple point, where all phases are at equilibrium.

\subsection{Droplet visualisation and comparison}

Despite having shown above that a finite-size droplet of the third phase wetting the interface between the two other phases is never the thermodynamic equilibrium state, due to the low miscibility of the oil and the water, it is possible to observe very long lived metastable states consisting of a droplet of one of the liquids at the interface between the other liquid and the vapour. Our (kitchen) experiments in Figs.~\ref{fig:kitchendrops} and \ref{fig:initialcomp}(a) clearly demonstrate this.
We now discuss how to calculate the density profiles for this sort of situation, such as that displayed in Fig.~\ref{fig:initialcomp}(b).

\begin{figure*}
    \centering
    \includegraphics[width=0.99\linewidth]{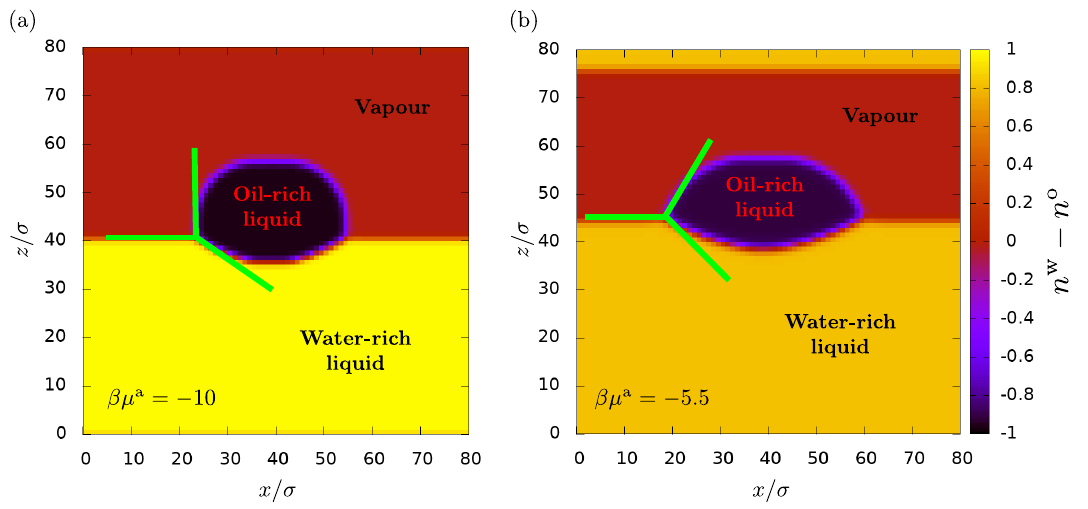}
    \caption{Equilibrated 2D oil-rich liquid droplets on a bath of water-rich liquid and the vapour above. Note the implementation of periodic boundary conditions (in both $x-$ and $z-$directions) leading to e.g.~the secondary interfaces between vapour and liquid near the top/bottom of the boxes. The green lines represent the Neumann angles, calculated independently and correspond to those found in Fig.~\ref{fig:thetsgams}. (a) is for the state point where $\beta\mu^{\textrm{a}}=-10$, with essentially no alcohol present in the system, and (b) for is $\beta\mu^{\textrm{a}}=-5.5$, which is the point in the phase diagram where there is roughly 16\% alcohol in the water-rich liquid phase and corresponding sizeable amounts all other phases. }
    \label{fig:drops}
\end{figure*}

To simplify our calculations, as mentioned above [c.f.\ Eq.~\eqref{eq:c_ij_liquids2D}], we assume that our system is invariant in one of the lattice directions and so is effectively 2D.
Using periodic boundary conditions on all sides of our box, we can again use Picard iteration to calculate the equilibrium density profiles.
We use the method described in Ref.~\onlinecite{archer2023stability}, here extended to include a third component.
The initial guess for the density profiles corresponds to setting the densities in the top half of the system equal to those of the vapour phase and in the bottom half corresponding to the coexisting water-rich liquid phase. Then, in a circular region in the centre of the box, we reset the density values to those of the oil-rich liquid phase.
We calculate the density profiles in a semi-grand canonical ensemble, in which the total number of oil molecules in the system $N^{\rm o}=\sum_\ii n_\ii^{\rm o}$ and also the total number of water molecules $N^{\rm w}=\sum_\ii n_\ii^{\rm w}$ are both fixed, while the amount of alcohol is allowed to vary, with the value of the chemical potential $\beta\mu^{\rm a}$ being fixed.
For further details on how these semi-grand canonical ensemble calculations are performed, see Refs.~\onlinecite{hughes2014introduction, archer2023stability}.
Examples of the resulting density profiles are displayed in Fig.~\ref{fig:drops}.
Due to the Laplace pressure contribution in such finite size oil droplets, the resulting (constrained) equilibrium system is no longer {\em exactly} at the triple point, i.e.\ the densities at the centre of the droplet and in the bulks of the water-rich liquid and vapour are slightly shifted from the bulk values at the triple point.
As the droplet is made larger, they do however approach the triple point values for the selected value of $\beta\mu^{\rm a}$.
Note also that for state points far away from the critical point, we can sometimes observe the interfaces between the different phases becoming weakly pinned to the underlying grid.
This issue was discussed in Ref.~\onlinecite{hughes2015liquid}, which also calculated the effective interaction (binding potential) between interfaces.
Here, we see this small effect manifest in the slightly flattened tops of the droplet profiles displayed in Fig.~\ref{fig:drops}.

The two equilibrated droplets shown in Fig.~\ref{fig:drops} correspond to the chemical potential values $\beta\mu^{\textrm{a}}=-10$ and $\beta\mu^{\textrm{a}}=-5.5$. These are two of the selected instances from our phase-diagram, corresponding to the corners of two of the `triangles' of tie-lines in Fig.~\ref{fig:TPhaseD}.
Figure~\ref{fig:drops} displays heatmap plots of the density difference $(n^{\textrm{w}}-n^{\textrm{o}})$, so that it is easy to identify the water-rich liquid phase having values near $+1$, the oil-rich liquid near $-1$, and vapour phase near zero.
Although this is slightly off-coexistence, our phase-diagram results at the same values of $\mu^{\textrm{a}}$ give us fairly closely the coexisting density values in each phase.
Then, from the results of Fig.~\ref{fig:thetsgams} we have the surface tensions and Neumann angles. To compare, we take the angles from the underlying computation of Fig.~\ref{fig:thetsgams}, where for $\beta\mu^{\textrm{a}}=-10$, we obtain the three angles $\theta_{\ell_\textrm{w}} = 145.4^\circ$, $\theta_\textrm{v} = 89.1^\circ$, $\theta_{\ell_\textrm{o}} = 125.5^\circ$ and 
for $\beta\mu^{\textrm{a}}=-5.5$, we obtain $\theta_{\ell_\textrm{w}} = 134.8^\circ$, $\theta_\textrm{v} = 120.8^\circ$, $\theta_{\ell_\textrm{o}} = 104.4^\circ$. In Fig.~\ref{fig:drops} we plot green lines on both droplets with these precise angles, seeing very good agreement---not withstanding the above remarks about the Laplace pressure shifting the system a little off from coexistence and the slight effects of grid pinning.

\section{Discussion and conclusion}
\label{sec:5}

In this paper we have predominately focused on extending our modelling approach of Ref.~\onlinecite{archer2024experimental} to uncover properties related to the coexisting vapour phase of ouzo. Before concluding, there are several aspects related to comparison with experiments worth noting first.

\subsection{Note on experimental comparison}

All results so far have been presented in a scaled (non-dimensional) form convenient for modelling.
To compare to experimental work, we often need to convert into corresponding dimensional quantities.
One set of quantities commonly dealt with in experiments are the mole fractions of the three components in any given phase.
In our model, we work with the three number densities $n^{\rm p}$, which vary between 0 and 1 and have the dimensions of number per unit volume.
However, recall that at the outset, for simplicity we set here the length scale of the underlying lattice $\sigma=1$ throughout, making the densities $n^{\rm p}$ effectively dimensionless.
The lattice site volume $\sigma^3$ can also be thought of as the average volume occupied by any of the molecules in the system \cite{archer2024experimental}.
The densities $n^{\rm p}$ do not contain information about the molecular weight of each species.
To compare with the experiments we need to convert these into mole fractions.
Due to our assumption that only one molecule of any of the three species can occupy a given lattice site at any given moment, we effectively assume that the volume occupied by all molecules are the same, which of course is not true in a physical system.
This means that one can define at least two different ways to convert number densities into mole fractions, which yield roughly the same answer, but are not fully consistent with each other, because of our modelling assumptions.

Firstly, for a given constituent $\textrm{p}$, where $\textrm{p}=\{\textrm{a},\textrm{o},\textrm{w}\}$, we have that
\begin{align}
    n^\textrm{p} = \frac{N^\textrm{p}}{V},
\end{align}
where $N^\textrm{p}$ is the number of molecules of the constituent $\textrm{p}$ in occupied volume $V$.
The number of moles is thus ${N^\textrm{p}}/{N_\textrm{A}}$, where $N_\textrm{A}$ is the Avogadro constant. Hence, one expression for the molar fractions is
\begin{align}
\label{eq:n_ratio}
    X^\textrm{p}(n) =
    \frac{{N^\textrm{p}}/{N_\textrm{A}}}{\Sigma_{\textrm{q}}({N^{\textrm{q}}}/{N_\textrm{A}})} = \frac{{N^\textrm{p}}}{\Sigma_{\textrm{q}}{N^{\textrm{q}}}}
    =\frac{{N^\textrm{p}}/{V}}{\Sigma_{\textrm{q}}({N^{\textrm{q}}}/{V})} = \frac{n^\textrm{p}}{\Sigma_{\textrm{q}}{n^{\textrm{q}}}},
\end{align}
where we have introduced the shorthand $n=\{n^\textrm{a},n^\textrm{o},n^\textrm{w}\}$. These are just the number density ratios for species p.

Alternatively, we can determine a total mass density based on the known mass densities of the pure liquids, $m_{\rm a}=0.79$ g/cm$^3$, $m_{\rm o}=0.99$ g/cm$^3$ and $m_{\rm w}=1.00$ g/cm$^3$, as done in Ref.~\onlinecite{archer2024experimental}. We can then follow the same process as would be done in experiments to obtain a mass fraction and then a molar fraction through the use of the molecular weights ($M^{\textrm{p}}$) of each species. These are: $M^{\textrm{w}} = 18$ for water, $M^{\textrm{a}} = 46$ for ethanol and $M^{\textrm{o}}=148$ for the trans-anethole. Therefore, the mass densities are given by 
\begin{align}
    \rho^\textrm{p} = n^\textrm{p}m^\textrm{p},
\end{align}
and the mass fractions by
\begin{align}
   w^\textrm{p} = \frac{\rho^\textrm{p}}{\rho},
\end{align}
where the total mass density $\rho = \Sigma_\textrm{p}{\rho^\textrm{p}}$.
From this, the molar fractions are then given by
\begin{align}
\nonumber
    X^\textrm{p}(w/M) 
    &= \frac{w^\textrm{p}/M^\textrm{p}}{\Sigma(w^{\textrm{q}}/M^{\textrm{q}})} 
    \\ &= \frac{\rho^\textrm{p}/M^\textrm{p}}{(\rho^\textrm{a}/M^\textrm{a})+ (\rho^\textrm{o}/M^\textrm{o}) + (\rho^\textrm{w}/M^\textrm{w})},
    \label{eq:molar}
\end{align}
where we have introduced the shorthand notation $w/M=\{w^\textrm{a}/M^\textrm{a},w^\textrm{o}/M^\textrm{o},w^\textrm{w}/M^\textrm{w}\}$.
If our model were exact, Eqs.~\eqref{eq:n_ratio} and \eqref{eq:molar} would give the same value.
However, they do not give quite the same, because of the lattice assumption discussed above.
We view the discrepancy between these two as a useful measure of the precision of our model.
In the following subsection we present some results illustrating this, however there is another aspect to all this that we should discuss first.

Recall that the values of the pair interaction strength parameters $\beta\epsilon^{\rm pq}$ used here are those given in Ref.~\onlinecite{archer2024experimental}. However, there are other possible sets of values that give a fair description of the ouzo system that one might consider.
As mentioned in Ref.~\onlinecite{archer2024experimental}, one can view this issue to be due to the fact that we are simplifying by mapping all the (complex multi-body) interactions of the real oil-water-alcohol system onto the simple lattice exclusions and pair interactions between neighbours.
The $\beta\epsilon^{\rm pq}$ values proposed in Ref.~\onlinecite{archer2024experimental} were identified based on trying to best match the overall shape of the room-temperature triangular ternary {\em liquid} phase diagram.
However, if other situations were considered, somewhat different values might be preferable. 
For example, changing the temperature (varying $\beta$) is straightforward, so we say very little about this here. At other temperatures, one can easily find cases where our theory predicts the phase-coexistence lines in the triangular ternary diagram meet the boundary, without the need of a critical point.
An interesting case to consider is just the pure ethanol-water mixture, because of the easy availability of literature data to compare with.

\subsection*{Ethanol-water system}
\label{sec:eth-wat}

Having developed our lattice model to describe the ternary ouzo system, it is illustrative to now apply it to the simpler (and much more widely investigated) miscible binary water-ethanol fluid system.
Since water and ethanol readily mix at all ratios, there are just two distinct fluid phases for this mixture, the liquid and the vapour.

\begin{figure*}
    (a) \hspace{5.5cm} (b) \hspace{5.5cm} (c) \hphantom{AAAAAAAAAAAAAAAAAAA}\\
    \centering
    \includegraphics[width=0.32\linewidth]{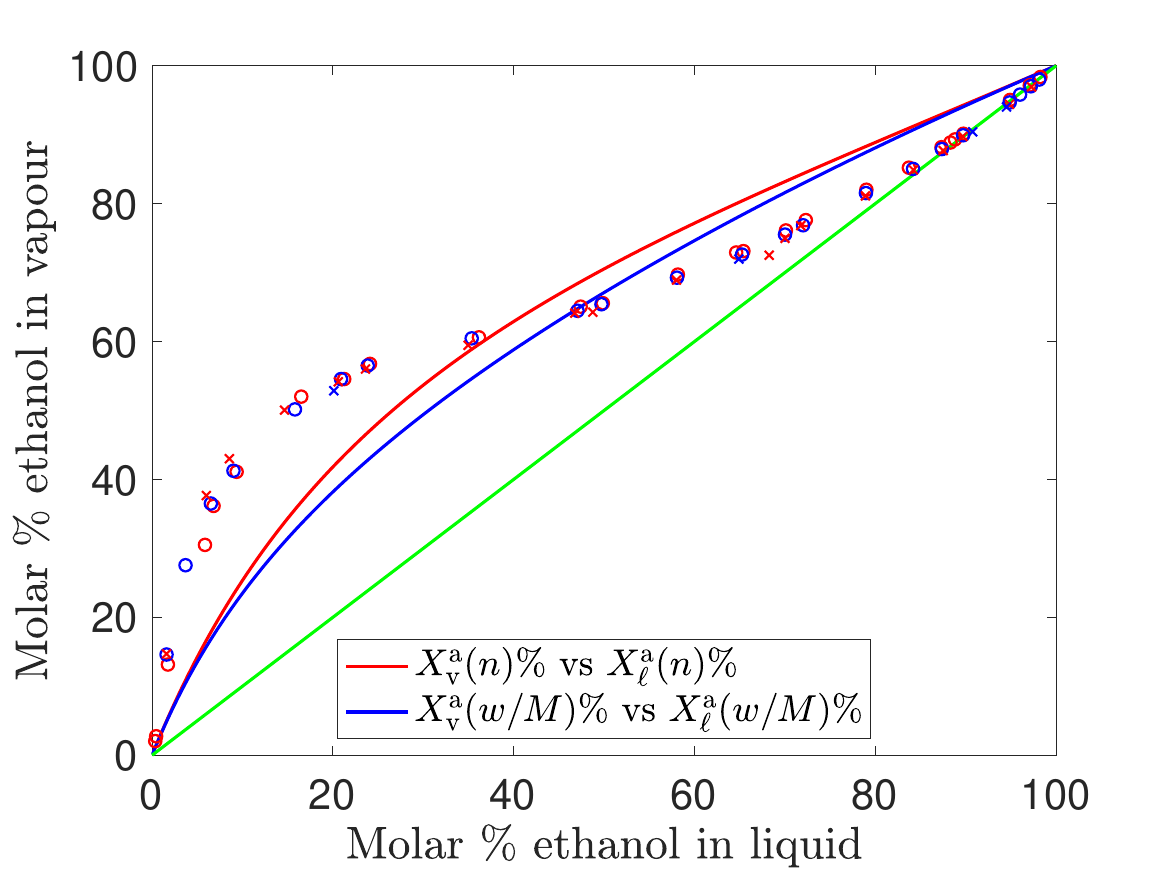} \,  \includegraphics[width=0.32\linewidth]{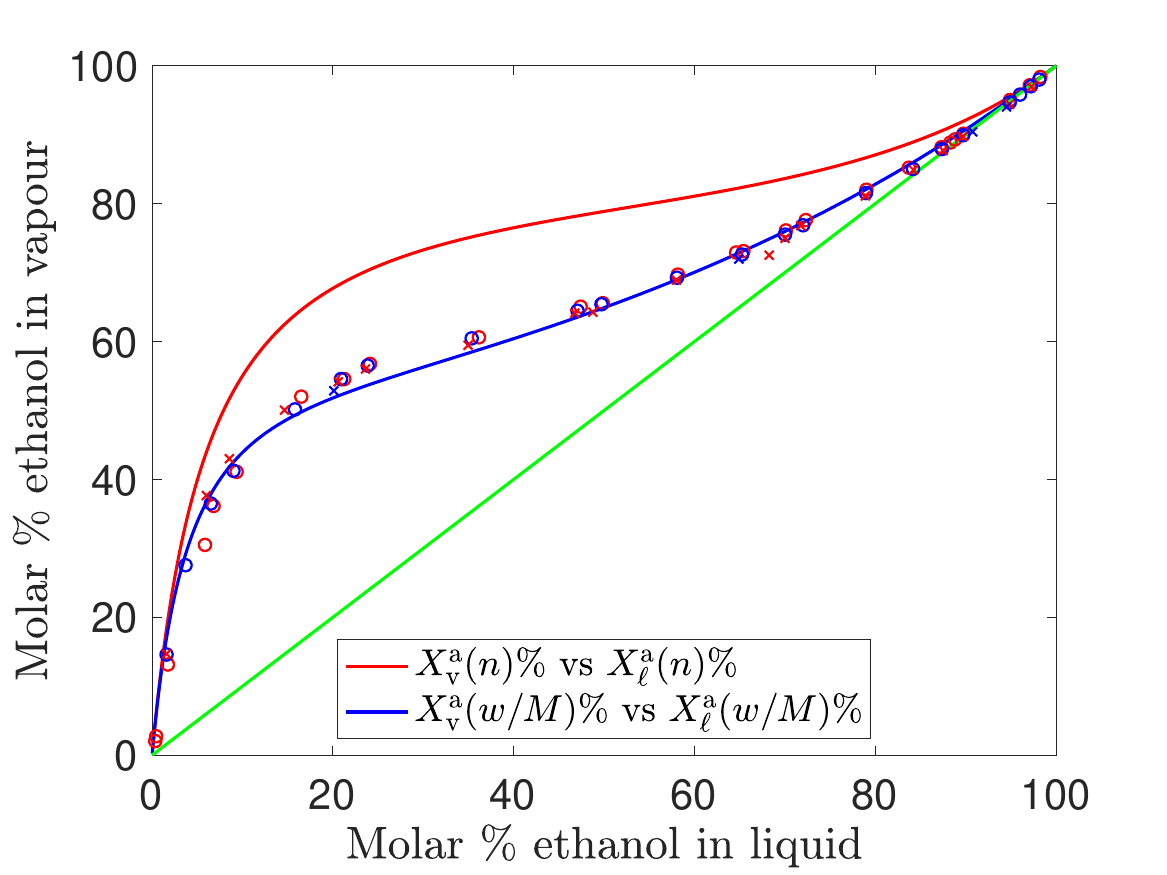} \,
    \includegraphics[width=0.32\linewidth]{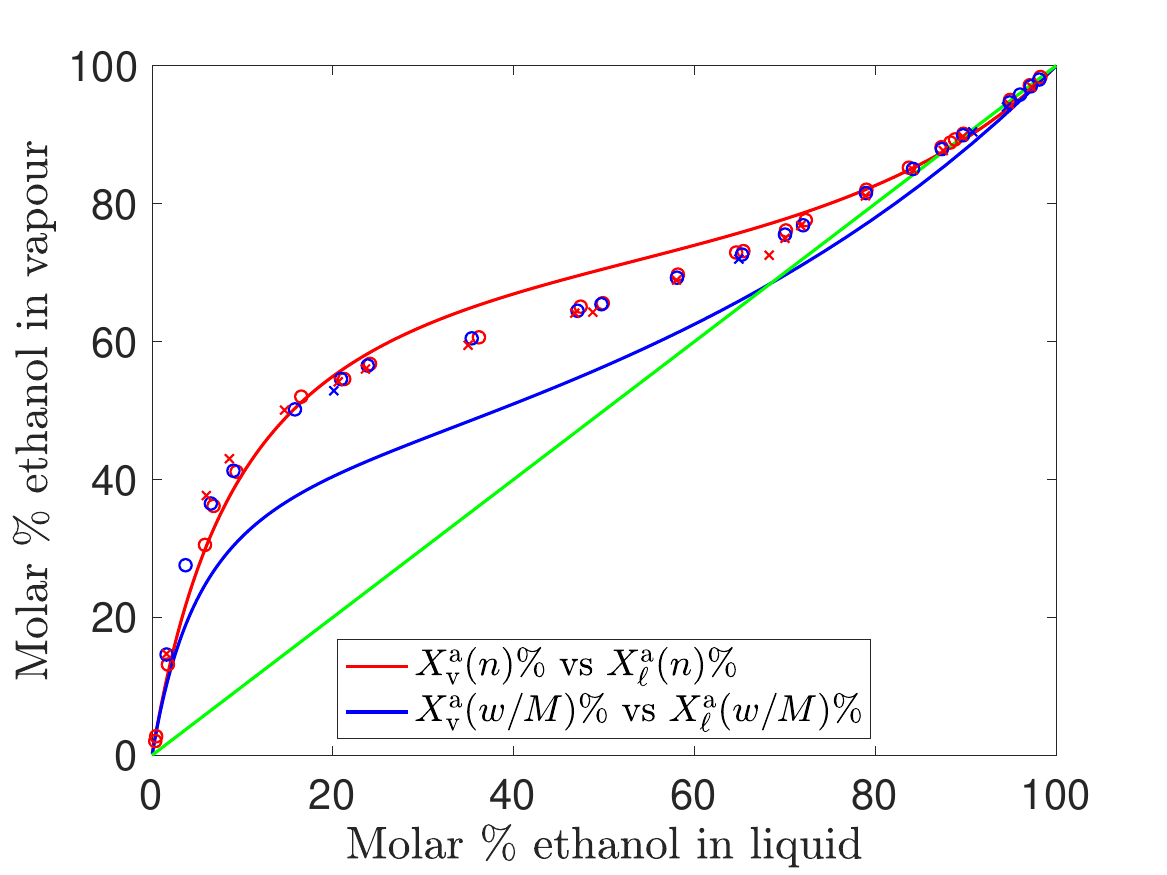}
    \caption{Comparison between experiments and our lattice model for the alcohol-water system with (a) the $\epsilon^\textrm{pq}$ values used for the ouzo system, i.e.~$\beta\epsilon_{\textrm{w}\textrm{w}} = 0.96$, $\beta\epsilon_{\textrm{a}\textrm{a}} = 0.78$, $\beta\epsilon_{\textrm{a}\textrm{w}} = 0.86$. (b) Changed values to fit the $(w/M)$-based molar fraction prediction better, with $\beta\epsilon_{\textrm{w}\textrm{w}} = 1.05$, $\beta\epsilon_{\textrm{a}\textrm{a}} = 0.78$, $\beta\epsilon_{\textrm{a}\textrm{w}} = 0.78$, and (c) again slightly changed values to fit the $n$-based molar fraction prediction, with $\beta\epsilon_{\textrm{w}\textrm{w}}=0.96$, $\beta\epsilon_{\textrm{a}\textrm{a}} =0.78$, $\beta\epsilon_{\textrm{a}\textrm{w}} = 0.75$. Symbols show experimental data from Ref.~\onlinecite{Beebe_etal}, with each symbol representing a different experimental pressure (distinguishing them is not relevant for the present work).}
    \label{fig:ethwatplot}
\end{figure*}

In Fig.~\ref{fig:ethwatplot} we plot the experimental data from Ref.~\onlinecite{Beebe_etal}, using their choice of representation of molar percentage of ethanol in the two possible phases (liquid and vapour), plotted against one another.
In Fig.~\ref{fig:ethwatplot}(a) we use the $\beta\epsilon^{\rm pq}$ values of the ouzo model discussed above, but setting $n^{\rm o}=0$, i.e.\ $\beta(\epsilon_{\textrm{w}\textrm{w}},\epsilon_{\textrm{a}\textrm{a}},\epsilon_{\textrm{a}\textrm{w}}) = (0.96,0.78,0.84)$, and plot the two possible molar fraction curves obtained via Eqs.~\eqref{eq:n_ratio} and \eqref{eq:molar}, for comparison.
We see that the model only qualitatively agrees with the experimental data, but correctly predicts that the curve bends up above the diagonal green line, indicating there is a higher fraction of ethanol in the vapour phase than in the coexisting liquid. However, neither the red nor the blue lines, calculated from Eqs.~\eqref{eq:n_ratio} and \eqref{eq:molar}, respectively, lie close to the experimental data (points).
Recall also our previous comments about the differences between calculating via Eqs.~\eqref{eq:n_ratio} and \eqref{eq:molar} being an indicator of the precision of our model.

An initial consideration of Fig.~\ref{fig:ethwatplot}(a) could be that our lattice model, only needing three parameters, $\epsilon^\textrm{ww}$, $\epsilon^\textrm{aw}$ and $\epsilon^\textrm{aa}$, for the pair interactions between the two components, is not sufficient to capturing the more nuanced real-world behaviour of the system.
However, making only small changes to one or two of the $\epsilon^\textrm{pq}$ values by eye, it is possible to achieve remarkably good agreement.
This is shown in Fig.~\ref{fig:ethwatplot}(b), where we change slightly the values of $\epsilon^\textrm{ww}$ and $\epsilon^\textrm{aw}$, so that $\beta(\epsilon_{\textrm{w}\textrm{w}},\epsilon_{\textrm{a}\textrm{a}},\epsilon_{\textrm{a}\textrm{w}}) = (1.05,0.78,0.78)$, which gives very good agreement with the $(w/M)$-based approximation of molar fraction $X^{\textrm{a}}$ in Eq.~\eqref{eq:molar} (blue line).
Another choice is to just change slightly the value of the cross interaction parameter $\epsilon^\textrm{aw}$, so that $\beta(\epsilon_{\textrm{w}\textrm{w}},\epsilon_{\textrm{a}\textrm{a}},\epsilon_{\textrm{a}\textrm{w}}) = (0.96,0.78,0.75)$, which is shown in Fig.~\ref{fig:ethwatplot}(c). This gives a decent fit to the Eq.~\eqref{eq:n_ratio} molar fraction prediction (red line).
In our view, what these three versions of our model provide is a useful measure of roughly what is the physically sensible range for the possible values of the $\epsilon^{\rm pq}$ parameters, i.e.\ roughly what are the `error-bars' on their values.
Recall that the $\epsilon^{\rm pq}$ values chosen in Ref.~\onlinecite{archer2024experimental} were selected to give good agreement for the full ternary ouzo system {\em liquid} phase diagram.
However, to apply the model more generally, we would consider potentially choosing slightly different values and would want to weigh up a variety of different experimental data (e.g.~in Ref.~\onlinecite{archer2024experimental} we put more emphasis on the overall asymmetry of the ternary phase diagram compared to trying to perfectly match the location of the critical point between model and experiment) and would aim to try and match best the data most relevant to the particular problem in hand.

\subsection*{Ouzo phase diagram}

\begin{figure}[t]
    \centering
    \includegraphics[width=0.98\linewidth]{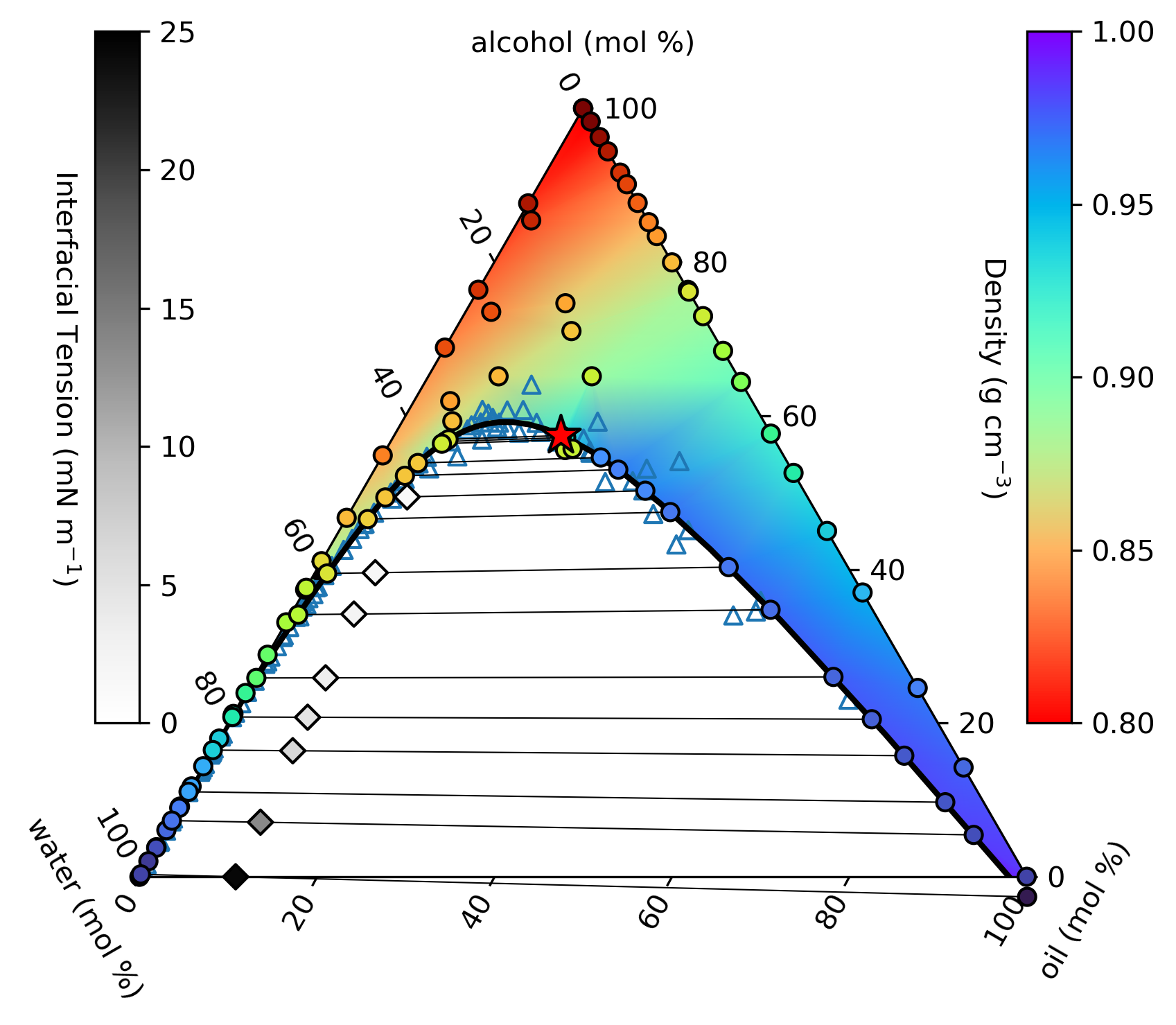}
    \caption{A revised ternary phase diagram using the same experimental data from Ref.~\onlinecite{archer2024experimental}, but plotting here using the respective molar fractions, instead of the mass fractions. Note that in this representation, the tie-lines appear very close to horizontal.}
    \label{fig:horiztie}
\end{figure}

In our previous work \cite{archer2024experimental}, we obtained the ouzo ternary phase diagram, in which we displayed  the coexisting state points (binodal points) in the standard way by plotting the mass fractions of the three constituents.
An interesting and intriguing observation is that if we
convert these to mole fractions (as discussed above), this yields the striking result that our experimental tie-lines become approximately horizontal. This is shown in Fig.~\ref{fig:horiztie}. 
This means that the significant asymmetry in molecular weights of our three ouzo constituents results in a rotation of points---notably the tie-lines and critical point---away from the heavier oil.
However, since our model predicts the critical point to be a little higher in the phase diagram than the experimental one, this same transformation rotates the model's tie-lines even further around, giving them a positive gradient (not displayed).

Whilst this observation is rather beautiful, we believe it is solely a feature specific to the ouzo system, and does not have any broader implications for the tie-lines in ternary systems in general.
In other words, there is something fortuitous about this particular transformation of the phase diagram, when it comes to the ouzo system.

\subsection{Concluding remarks}

Water, food oils and alcoholic spirits like ouzo are all liquids that can be used in simple home experiments to gain a hands-on understanding of various interfacial phenomena, including interfacial tensions, the behaviour of droplets at interfaces and the resulting Neumann angles.
Ouzo is a particularly fascinating liquid, because of the dissolved trans-anethole and the spontaneous emulsion formation that results from just adding water to it.
By performing these experiments in conjunction with the calculations presented here that are based on a simple lattice DFT, we have shown how to gain an even deeper understanding of this physics of fluids.
We have shown that our DFT is able to reliably predict quantities like droplet shapes and Neumann angles.
Given the basic ingredients in our lattice model and the minimal level of information regarding aspects like the molecular correlations in this ternary liquid mixture, the simple DFT fares remarkably well.
Thus, although the present work has been presented in the spirit of kitchen science, it is also valuable in paving the way for applying our DFT approach more widely to the bulk and interfacial phase behaviour of multicomponent fluid mixtures of the type discussed here.

The work presented here has shown that our simple model is capable of describing the vapour phase of the ternary ouzo mixture and also the simpler water-alcohol binary system.
Our focus here has predominantly been on three-phase coexistence (triple point) equilibrium between oil-rich and water-rich liquids and the vapour phase, where all three phases contain the constituents alcohol (ethanol), oil (trans-anethole) and water.
However, we are confident that the model can easily be applied (maybe with some minor adaptations) to other thermodynamic state points where the system is still fluid.
One possible future direction is to improve the accuracy of the DFT by using lattice fundamental measure theory \cite{lafuente2002fundamental, lafuente2004density, lafuente2005cluster, maeritz2021density, maeritz2021droplet, zimmermann2024lattice}. 

Whilst presenting this, we have called into question the best way to visualise the phase behaviour, particularly due to the classic ternary phase diagram being a projection that loses information about overall density, which is especially important here given that the vapour phase is generally of significantly lower density than the liquid phases.

Our work has probed the liquid interfacial behaviour between phases, including the profiles of the constituents, and then computing emergent properties such as surface tensions and Neumann angles between phases.
Given this success, we expect our model to be reasonably accurate and easy to apply to determine the wetting behaviour of ouzo and its mixtures at solid surfaces.
Indeed, a two-component version of the model has already been used (albeit with different values of $\epsilon^{\rm pq}$), to elucidate the behaviour of binary liquids at surfaces and to calculate the binding potential (i.e.\ the effective interaction between interfaces) for binary mixtures on solid surfaces \cite{areshi2024binding}.

Whilst initially we conceived of the present work as an extension of our earlier work on ouzo \cite{archer2024experimental}, the richness of the behaviour occurring with the vapour phase alongside the liquids means a number of future research directions would be very interesting. One area is the modelling of interfacial behaviour in dynamic situations, such as the growth of water-rich droplets in a reservoir of ouzo\cite{ouzovideo}.
An interesting modelling approach would be to develop an effective interface Hamiltonian based model, similar to those used for colloidal fluids \cite{Zhang_Sibley_Tseluiko_Archer_2024} or ice-water-vapour systems \cite{sibley2021ice}.
Such models enable us to easily incorporate interface motion through mass transfer such as via evaporation/condensation to/from the vapour phase into the modelling.
Effective binding potentials between interfaces of the phases could be obtained from the present lattice density functional theory using a procedure similar to that implemented for simple liquids \cite{hughes2015liquid, areshi2024binding} and bubbles \cite{Yin_2019}.

Another future research area is a much more thorough study of phase stability, and particularly the dynamics that can occur.
This can straightforwardly be done by extending the dynamical DFT approach developed in Ref.~\onlinecite{chalmers2017dynamical} to ternary mixtures.
Our spinodal computations suggest that some regions of phase space where vapour bubbles should cavitate in either liquid phase are fairly easy to access, as well as the more well-understood liquid-liquid spinodal decomposition that leads to the easily observed ouzo effect.
Experiments\cite{Vratsanos} and dynamical modelling of this would be fascinating. Models based on hydrodynamic density-functional theory exist \cite{archer2009dynamical, hydrodynamicDDFT} for simple fluids, that are able to resolve microscopic interfacial and contact-line properties, so in conjunction with the present reliable theory for the interfaces of alcohol, water and oil mixtures, such a dynamical extension of the theory would also enable studies of the evaporation properties (see also Ref.~\onlinecite{chalmers2017dynamical}), giving e.g.\ the evolving dynamical properties of the head-space of alcoholic beverages, an important factor in the flavour experience \cite{jurado2007characterization, pohjanheimo2009headspace}.

\section*{Author Declarations}
The authors have no conflicts to disclose.

\section*{Data Availability}
The data that support the findings of this study are available from the corresponding author upon reasonable request.

\section*{Acknowledgements}

We are grateful to Nathan Bousquet, James Swift, James Reynolds and Matthew Turner for valuable discussions and to Iain Phillips for technical support. 
The contributions of A.J.A.\ and B.D.G.\ to this project were partially supported by the London Mathematical Society and the Loughborough University Institute of Advanced Studies.

%\bibliographystyle{abbrv}
%\bibliography{refs}

\end{document}